\documentclass[journal]{IEEEtran}
\IEEEoverridecommandlockouts
\usepackage{cite}
\usepackage{amsmath,amssymb,amsfonts}
\usepackage{algorithmic}
\usepackage{algorithm}
\usepackage{graphicx}
\usepackage{textcomp}
\usepackage{xcolor}
\usepackage{pifont}
\usepackage{cases}
\usepackage{url}
\usepackage{multirow}
\usepackage{bm}
\usepackage{mathrsfs}
\newtheorem{definition}{ \textbf{Definition}}

\newcommand{\tabincell}[2]{\begin{tabular}{@{}#1@{}}#2\end{tabular}}
\usepackage{eqparbox}

\usepackage{geometry}
\begin{document}

\title{A Secure and Intelligent Data Sharing Scheme for UAV-Assisted Disaster Rescue}

\author{Yuntao~Wang, Zhou~Su, Qichao~Xu, Ruidong~Li, Tom~H.~Luan, and Pinghui~Wang
\thanks{Y. Wang, Z. Su, and T. H. Luan are with the School of Cyber Science and Engineering, Xi'an Jiaotong University, Xi'an, China (\emph{Corresponding author: Zhou~Su}).}
\thanks{Q. Xu is with the School of Mechatronic Engineering and Automation, Shanghai University, China.}
\thanks{R. Li is with the Department of Electrical and Computer Engineering, Kanazawa University, Japan.}
\thanks{P. Wang is with MOE KLINNS Laboratory, Xi'an Jiaotong University, China}}

\maketitle

\begin{abstract}
Unmanned aerial vehicles (UAVs) {have the potential to} establish flexible and reliable emergency networks in disaster sites when terrestrial communication infrastructures go down.
Nevertheless, potential security threats may occur on UAVs during data transmissions due to the {untrusted} environment and open-access UAV networks. Moreover, UAVs typically have limited battery and computation capacity, making them unaffordable for heavy security provisioning operations when performing complicated rescue tasks.
In this paper, we develop RescueChain, a secure and efficient information sharing scheme for UAV-assisted disaster rescue.
Specifically, we first implement a lightweight blockchain-based framework to safeguard data sharing under disasters and immutably trace misbehaving entities.
A reputation-based consensus protocol is devised to adapt the weakly connected environment with improved consensus efficiency and promoted UAVs' honest behaviors.
Furthermore, we introduce a novel vehicular fog computing (VFC)-based off-chain mechanism by leveraging ground vehicles as moving fog nodes to offload UAVs' heavy data processing and storage tasks.
{To offload computational tasks from the UAVs to ground vehicles having idle computing resources, an optimal allocation strategy is developed by choosing payoffs that achieve equilibrium in a Stackelberg game formulation of the allocation problem.}
For lack of sufficient knowledge on network model parameters and users' private cost parameters in practical environment, we also design a two-tier deep reinforcement learning-based algorithm to seek the optimal payment and resource strategies of UAVs and vehicles with improved learning efficiency.
Simulation results show that RescueChain can effectively accelerate consensus process, improve offloading efficiency, reduce energy consumption, and enhance user payoffs.
\end{abstract}

\section{Introduction}
Natural disasters, such as earthquakes, bushfires, floods, often inflict devastating losses in lives and property \cite{Wang2021INFOCOM,9252867,9696188}. In disaster areas, reliable and resilient emergency communications are pivotal for quick damage assessment and effective disaster rescue \cite{9035635}.
However, owing to the destruction or inefficiency of terrestrial network infrastructures (e.g., Wi-Fi access points and cellular base stations),
unmanned aerial vehicles (UAVs) based communication approach may be the only plausible solution in such situations so far due to their fast deployment and flexible mobility \cite{ToN9219140,Liu2019tran,ToN9003500}. UAVs can not only be dispatched and deployed quickly to establish the emergency communication infrastructure on the air, but also help information dissemination (e.g., rescue commands, maps of affected areas, and survivors' locations) in disaster relief networks \cite{9599638,8468999,7807176}.

There are however several fundamental challenges underlying the UAV-aided disaster relief networks (UDRNs). Since UAVs need to be open to all nodes, including malicious entities, in the disaster area to maximize the life-saving potential, the network is vulnerable to various attacks (e.g., spoofing attack and DDoS attack) perpetrated by adversaries \cite{9035635}. The UAVs may also be compromised to forge, remove, and replace the transmitted data, and even inject malwares and viruses to mislead or interfere with the rescue operations.
In addition, the onboard battery and computing capabilities of UAVs are typically limited, whereas the heavy computation missions in affected areas such as human detection and video recognition often exceed UAVs' local processing capabilities \cite{9155397,Erdelj2017Wireless}. Consequently, the time required to execute search and rescue operations may be overextended and the efficiency of disaster rescue may be degraded.
Therefore, a secure data transmission scheme with efficient computation and storage offloading should be devised for UAVs in UDRNs.

The emerging blockchain holds numerous potentials to build trust among various collaborative entities; by offering decentralized hash-linked ledgers with time-stamped data and behavior records, the blockchain prevents fraud in a reliable and distributed manner \cite{9631953,ToN9036064,9928220}. Participants can share and retrieve the desired data on the basis of blockchain which is featured with immutability, transparency, and auditability. For example, in \cite{Islam2019buav}, a decentralized data collection mechanism is designed based on blockchain to safeguard data delivery in UAV-assisted IoT. In \cite{Li2019block}, a blockchain-based group key distribution mechanism is developed to build trust among UAVs and ensure the security of UAVs' sensory data. 

Existing blockchain approaches for UAVs heavily depend on the availability of infrastructures for security-critical operations (e.g., consensus management and ledger maintenance). A fully distributed approach tolerable to UAV failures is more suitable for practical disaster scenarios. Meanwhile, the compute-intensive data processing and consensus operations along with heavy storage requirements in blockchain-enabled UDRNs often occupy a large amount of constrained computation and storage for UAVs.
Existing works mainly focus on cloud or edge based approaches for UAVs' computation and storage offloading \cite{Bai2019edge,8654698,9093980}; nonetheless, due to the long distance of remote clouds and the insufficiency or unavailability of edge servers under disasters, the stringent quality of service (QoS) requirements of UAVs (e.g., completing missions before expiration) may fail to meet in the current cloud or edge based offloading approaches.
Hence, it is still an open and vital issue to secure data sharing by deploying a lightweight and robust blockchain system in disaster areas while efficiently offloading UAVs' heavy computation and storage tasks.

In this paper, we develop RescueChain, a novel secure and energy-efficient data sharing scheme for UDRNs. We first propose a lightweight and infrastructure-free blockchain-based framework to safeguard data sharing {and immutably trace misbehaving entities} in disaster sites. We then devise a reputation-based Tendermint consensus protocol to efficiently and robustly reach consensus under weak network connections while encouraging UAVs' legitimate behaviors in the network. 
Afterwards, by exploiting ground vehicles as moving fog nodes, a novel vehicular fog computing (VFC)-based off-chain computation and storage mechanism is presented to collaboratively offload UAVs' data processing and security provisioning missions to moving vehicles.
The Stackelberg game model is formulated to model the interactions between UAVs and vehicles under VFC and the Stackelberg equilibrium (SE) of the static Stackelberg game is derived to stimulate vehicles' participation in computing resources sharing.
As accurate network parameters and the private user cost model are not readily available in realistic offloading applications, a dynamic Stackelberg game is presented, and its near-optimal solution is sought by using an intelligent learning algorithm based on deep reinforcement learning (DRL) techniques.
The contribution of this paper is three-fold as follows.
\begin{itemize}
\item \emph{Framework:} We present RescueChain, which is \emph{low-cost}, \emph{infrastructure-free}, and \emph{robust}. We make two improvements in the implementation of blockchain in UDRNs: 1) a green consensus protocol with non-mining bookkeeping, low communication complexity, and high robustness under partially synchronous environment; 2) a VFC-based off-chain data storage and computation mechanism to efficiently move the heavy data processing and storage missions from resource-constrained UAVs to nearby cooperative ground vehicles.
\item \emph{Algorithm:} Due to sparse network connections in UDRNs, we develop an energy-efficient and partition-tolerant consensus algorithm for battery-limited UAVs based on Tendermint via signature aggregation and reputation evaluation. {Besides, nodes' benign and malicious behaviors can be immutably traced and recorded on the redesigned blockchain ledgers for reputation computing.}
    To promote vehicles' collaboration under VFC in the fast-changing environment, a learning-based algorithm in the dynamic Stackelberg game is designed to intelligently schedule the optimal resource sharing and pricing strategies for ground vehicles and UAVs.
    To cope with the large space size and address the curse of dimensionality in learning, we also exploit DRL techniques with two tiers for efficient state space compression and accelerated convergence rate.
\item \emph{Validation:} We evaluate the effectiveness of RescueChain through extensive simulations. It is demonstrated that our RescueChain can attain better payoffs for ground vehicles and UAVs, reduced data delivery latency and UAVs' energy consumption, improved offloading efficiency, and enhanced consensus efficiency in blockchain, by comparing with other existing schemes.
\end{itemize}

\begin{figure}[t!]\setlength{\abovecaptionskip}{-0.05cm}
\centering
\setlength{\abovecaptionskip}{-0.25cm}
  \includegraphics[width=8cm]{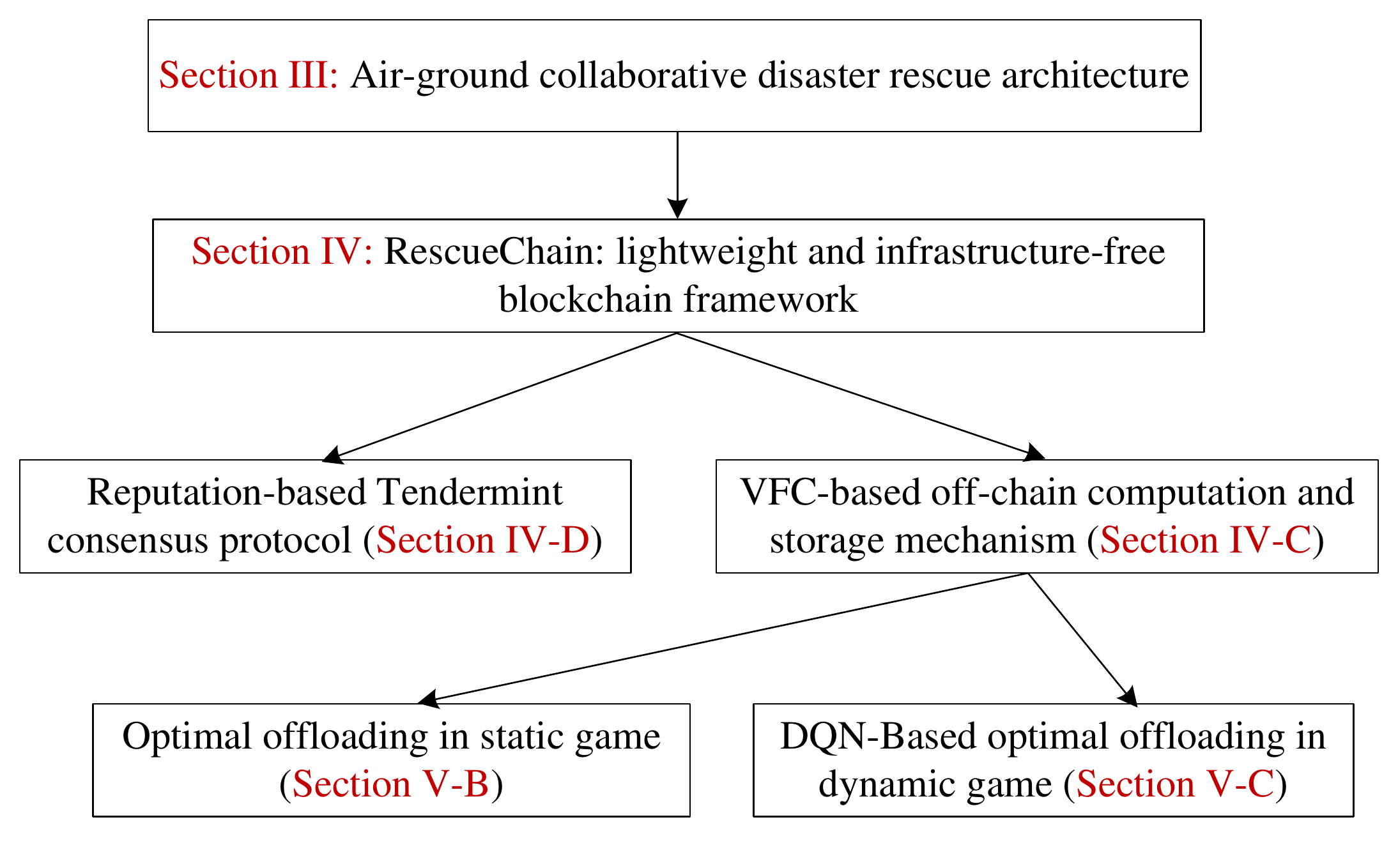}
  \caption{{Main research contents and their organization structure.}}\label{fig:organization}\vspace{-0.24cm}
\end{figure}

The remainder of the work is organized as follows. Related works are reviewed in Section \ref{sec:RELATED WORK}. The system model is elaborated in Section \ref{sec:SYSTEM MODEL}. The design of RescueChain system is presented in Section \ref{sec:SOLUTION1}. The optimal offloading policies in static and dynamic games are given in Section \ref{sec:SOLUTION2}. Section \ref{sec:SIMULATION} evaluates the proposed scheme and Section \ref{sec:CONSLUSION} closes this paper with conclusions. {For reader's convenience, the main research contents and their organization structure are illustrated in Fig.~\ref{fig:organization}.}

\section{Related Work}\label{sec:RELATED WORK}
In this section, we review the related works on blockchain-based information sharing approaches and computation offloading methods for UAVs.
\vspace{-0.3cm}
\subsection{Blockchain-Based Data Sharing}
The emerging blockchain technologies in wireless networks for secure data sharing have attracted wide attention from both academia and industry.
Liang \emph{et al}. \cite{8673633} design a permissioned blockchain-based electricity trading platform in industrial Internet of things (IoT) to address the high management cost, ensure reliable data transmission, and build trust among grid operators and electricity consumers.
By leveraging the consortium blockchain, Chen \emph{et al}. \cite{8863960} develop a trustful on-chain and off-chain data sharing framework in vehicular networks, where edge nodes (i.e., roadside units (RSUs)) are responsible for blockchain maintenance by operating the practical Byzantine fault tolerance (PBFT) consensus protocol.
Jiang \emph{et al}. \cite{8963705} investigate a blockchain-based reliable artificial intelligence (AI) model sharing mechanism for object detection in autonomous driving with cross-domain adaptation, where mobile edge computing (MEC) nodes run the delegated proof-of-stake (DPoS) algorithm for efficient ledger management.
By deploying a vehicular blockchain network with homomorphic cryptosystem, Kong \emph{et al}. \cite{9069253} propose a secure and verifiable sensory information collection and sharing scheme in fog computing-enabled IoV, where the consensus process in blockchain is managed by RSUs with PBFT protocol.
One can observe that most of the existing blockchain systems heavily depend on communication infrastructures for block building and consensus management and cannot be directly applied for disaster areas with severely disrupted or unavailable infrastructures.
\vspace{-0.3cm}
\subsection{Computation Offloading for UAVs}
Recently, many works have been reported on computation offloading for UAVs with constrained onboard resources.
Bai \emph{et al}. \cite{Bai2019edge} design an MEC-based energy-efficient task offloading framework for UAVs in wireless networks with consideration of active and passive eavesdroppers and time-duration constraints.
By exploiting the edge infrastructures, Callegaro \emph{et al}. \cite{8648099} propose an edge computing-based computation offloading scheme and design the optimal offloading policies for UAVs with joint consideration of network and computation load of MEC nodes.
Chen \emph{et al}. \cite{Chen2019cloud} present a hybrid edge/cloud computing-based offloading scheme to efficiently offload UAVs' real-time computation tasks to proximal edge servers or remote cloud servers to reduce latency and improve energy efficiency.
Liu \emph{et al}. \cite{9093980} devise an online multi-hop trajectory scheduling and computation task assignment algorithm for UAVs based on Lyapunov optimization and Markov approximation approaches for joint computing delay and cost minimization in the edge-cloud environment.
However, as network infrastructures can be unavailable or insufficient in disaster scenarios, conventional cloud or edge based offloading mechanisms are difficult to meet the QoS requirements of UAVs.
Moreover, different from the assumption in most of the existing works, it is not readily available for both resource consumers and resource contributors to acquire the accurate cost model parameters to determine the optimal offloading policies in practice.

In this work, distinguished from existing works, we study a lightweight and green consensus protocol for disaster scenarios with weak network connections to facilitate the implementation of blockchain into UDRNs.
Besides, we devise an intelligent DRL-based optimal offloading algorithm under the promising VFC paradigm by exploiting the idle computation and storage resources shared by ground vehicles.
Compared with our previous work \cite{Wang2021INFOCOM}, new contributions of this paper include:
(1) we design a misbehavior tracing mechanism and give the property analysis of our RescueChain system;
(2) we derive the SE for the general case of VFC-based offloading with multiple UAVs and vehicles with continuous strategies; and
(3) we develop a two-tier deep Q-network (DQN)-based offloading mechanism to enhance the performance of Q-learning based offloading mechanism in \cite{Wang2021INFOCOM}.

\section{System Model}\label{sec:SYSTEM MODEL}
In this section, we introduce the system model including the network model, mobility model, VFC model, communication model, and adversary model, respectively. A summary of notations used in the paper is presented in Table \ref{table1}.

\begin{figure}[t!]\setlength{\abovecaptionskip}{-0.0cm}
\centering
  \includegraphics[width=8.3cm,height=5.05cm]{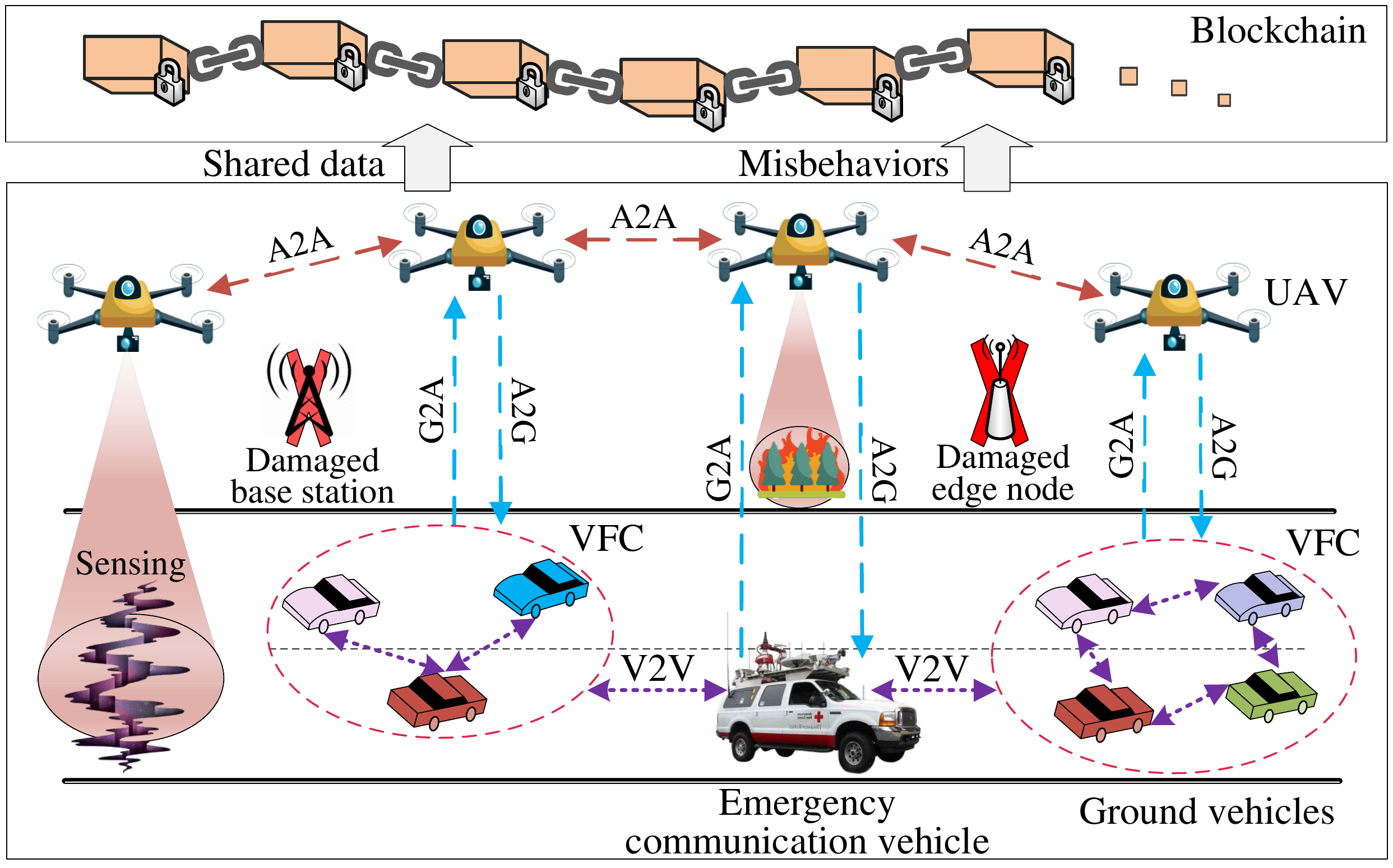}
  \caption{System model of RescueChain in a disaster area \cite{Wang2021INFOCOM}.}\label{fig:model}\vspace{-0.2cm}
\end{figure}

\begin{table*}\scriptsize
\setlength{\abovecaptionskip}{-0.04cm}
\caption{Summary of Notations}\label{table1}\centering
\begin{tabular}{|c|l||c|l|}
\hline
\textbf{Notation} & \textbf{Description} & \textbf{Notation} & \textbf{Description} \\  \hline 
$\mathcal{I}$&Set of ground vehicles. &$Z$&Number of validators in the blockchain. \\
$\mathcal{J}$&Set of UAVs. &$\Psi$&Number of level-1 validators. \\
${\mathcal{I}_j}[t_n] $&Set of vehicles at UAV $j$'s coverage at $n$-th time slot. &$\mathcal{T}_1$&Merkle tree of time-stamped transactions in the block body. \\
$\mathcal{Z}$&Set of elected validators in the blockchain. &$\mathcal{T}_2$&Merkle tree of serialized chunks of the raw block. \\
$\mathcal{U}$&Set of full nodes in the blockchain. &$\mathcal{T}_3$&Merkle tree of signed evidences for misbehavior forensics. \\
$N$&Total number of time slots. &$r,h$&Round/height in the consensus process. \\
$R_j^{A2G},R_j^{A2A}$&Radius of A2G/A2A communication range of UAV $j$. &$\hbar$&The leader designated from level-1 validators. \\
$\mathbf{l}_j[t_n]$&Horizontal location of UAV $j$ at $n$-th time slot. &$\Upsilon_{\hbar}$&Proof-of-lock (PoL) of the leader ${\hbar}$. \\
$H_j$&Flying altitude of UAV $j$. &$\sigma_{\hbar},\sigma_{z}$&Signature of leader ${\hbar}$/validator $z$. \\
$v_j[t_n],v_j^{\max}$&Velocity/maximum velocity of UAV $j$ at $n$-th time slot. &$tx_{\mathrm{data}}$&Off-chain storage transaction. \\
$a_j[t_n]$&Acceleration of UAV $j$ at $n$-th time slot. &$tx_{\mathrm{rep}}$&Report transaction of nodes' misbehaviors. \\
$\overline v_{GV}$&Average vehicular velocity. &$LastProof$&A script for misbehavior forensics. \\
${v_{GV}^{\max }},{v_{GV}^{\min }}$&Maximum/minimum vehicle velocity. &$LastCommit$&Precommits making the prior block at height $h$ committed. \\
$\chi,{\chi _{\max }}$&Traffic density/maximum traffic density of vehicles. &$LastVotes$&Votes justifying precommits at all rounds for height $h$. \\
$K_j$&Total number of UAV $j$'s tasks to be offloaded. &$\Im_u^{n}$&Reputation value of full node $u$. \\[1pt]
$\Gamma_{j,k}$&$k$-th computation task offloaded by UAV $j$. &${{\overline{\Im}_u^{n}}}$&Normalized reputation value of full node $u$. \\
$D_{j,k}$&Data size of task $\Gamma_{j,k}$. &$\Im_u^{\mathrm{ini}}$&Initial reputation value of full node $u$. \\
$\theta_{i,j,k}$&Required CPU cycles to process one bit of task $\Gamma_{j,k}$. &$\Delta_b$&Reputation reward/penalty for behavior $b$. \\
$T_{j,k}^{\max}$&Time-to-live (TTL) of task $\Gamma_{j,k}$. &${x_{i,j,k}}$&Amount of computing resource (AoCR) in doing task $\Gamma_{j,k}$. \\
$\alpha_{j,k}$&Urgency degree of task $\Gamma_{j,k}$. &${y_{i,j,k}}$&Payment of task $\Gamma_{j,k}$. \\
$\mu_{j,k}$&Output/input ratio of task $\Gamma_{j,k}$. &$S_{j,k}(x_{i,j,k})$&Satisfaction function of UAV $j$. \\
$E_j[t_n]$&UAV $j$'s remaining battery energy at $n$-th time slot. &$C_{j,k}(x_{i,j,k},y_{i,j,k})$&Cost function of UAV $j$. \\
$E_j^{\min}$&Minimum energy reserve of UAV $j$. &$\pi _j\left(\mathbf{x}_j, \mathbf{y}_j\right)$&Payoff function of UAV $j$. \\
$C_j$&Battery energy capacity of UAV $j$. &${\pi _i}({x_{i,j,k}},y_{i,j,k})$&Payoff function of ground vehicle $i$. \\
$P_j^{\mathrm{fly}}[t_n]$&Flying or propulsion power of UAV $j$ at $n$-th time slot. &$\lambda_p,\lambda_c,\lambda_e$&Positive adjustment parameters. \\[2pt]
$P_i^{\mathrm{TX}},P_j^{\mathrm{TX}}$&Transmit power of vehicle $i$/UAV $j$. &$\Psi_i(x_{i,j,k})$&Cost function of vehicle $i$. \\
$E_{i,j,k}^{\mathrm{vfc}}$&Energy consumption of vehicle $i$ in task execution. &$\mathbf{s}_j^{(n)}$&State of UAV $j$ at time slot $n$. \\[2pt]
$E_{i,j,k}^{\mathrm{fly}}$&Energy consumption of UAV $j$ in movement. &${\tilde{s}}_{i,j,k}^{(n)}$&State of vehicle $i$ at time slot $n$. \\
$E_{i,j,k}^{\mathrm{A2G}}$&Energy consumption of UAV $j$ in A2G transmission. &$Q(\mathbf{s}_j^{(n)},\mathbf{y}_j^{(n)} )$&Q-function of UAV $j$ for state-action pair $(\mathbf{s}_j^{(n)},\mathbf{y}_j^{(n)})$. \\
$\varphi _{i,j}$&Channel gain between UAV $j$ and vehicle $i$. &$\tilde{Q}({\tilde{s}}_{i,j,k}^{(n)},{x}_{i,j,k}^{(n)})$&Q-function of vehicle $i$ for state-action pair $({\tilde{s}}_{i,j,k}^{(n)},{x}_{i,j,k}^{(n)})$. \\
$d_{i,j}[t_n]$&Euclidean distance between UAV $j$ and vehicle $i$. &$\varsigma _j,\varsigma _i$&Discount factor of UAV $j$/vehicle $i$. \\
${\gamma _{i,j}^{\mathrm{G2A}}},{\gamma _{i,j}^{\mathrm{A2G}}}$&Transmission rate of G2A uplink/A2G downlink. &${\bm{\gamma}}_j^{(n)},{\bm{\gamma}}_i^{(n)}$&State sequence of UAV $j$/vehicle $i$ in DQN. \\
$B_{i}^{\mathrm{UL}},B_{j}^{\mathrm{DL}}$&Uplink/downlink bandwidth. &${\bm{\phi}}_j^{(n)},{\bm{\phi}}_i^{(n)}$&Interaction experience of UAV $j$/vehicle $i$ in DQN. \\
$\Phi_{i,j,k}$&Total offloading latency of task $\Gamma_{j,k}$. &$\Xi_j,\Xi_i$&Replay memory of UAV $j$/vehicle $i$ in DQN. \\[1pt]
$t_{i,j,k}^{\mathrm{vfc}}$&Execution time of task $\Gamma_{j,k}$. &$\epsilon_j,\epsilon_i$&Parameter of $\epsilon$-greedy policy of UAV $j$/vehicle $i$. \\[2.3pt]
$t_{i,j,k}^{\mathrm{A2G}}$&A2G transmission time of task $\Gamma_{j,k}$. &$W$&Number of UAV's payment levels in DQN. \\[2.3pt]
$t_{i,j,k}^{\mathrm{G2A}}$&G2A transmission time of task $\Gamma_{j,k}$. &$V$&Number of vehicle's AoCR levels in DQN. \\[1.8pt]
\hline
\end{tabular} 
\end{table*}

\vspace{-0.12cm}
\subsection{Network Model}
Fig.~\ref{fig:model} depicts a typical UAV-aided disaster rescue network (UDRN), which is composed of a group of UAVs, ground vehicles, ground stations, and a permissioned blockchain. More basic information about UDRNs can be found in works \cite{7807176,9599638}. 

In a given investigated disaster area, a set of vehicles, denoted as $\mathcal{I} = \{1,\cdots,i,\cdots,I\}$, are to perform search and rescue missions, where part of roads and network infrastructures are damaged. To facilitate data transmissions among ground rescue vehicles, a fleet of UAVs, denoted as $\mathcal{J} = \{1,\cdots,j,\cdots,J\}$, are dispatched to establish emergency communications by forming flexible aerial subnetworks.
Each UAV $j\in \mathcal{J}$ is equipped with multiple sensors (e.g., thermometer, infrared camera, and GPS) to perceive its surroundings.
On one hand, to save UAV's limited battery energy, the sensory data (e.g., images, videos, and audios) collected by UAVs can be transmitted to ground vehicles for processing (referred to as VFC) such as pattern recognition and survivor detection via aerial-to-ground (A2G) links.
On the other hand, a group of UAVs can serve as aerial communication relays via aerial-to-aerial (A2A) links to enhance network connectivity. Let $R_j^{A2G}$ and $R_j^{A2A}$ denote the radiuses of A2G and A2A communication ranges of UAV $j$, respectively. Due to the destruction of network infrastructures, the emergency control centers such as ground stations are deployed to coordinate both aerial and ground subnetworks by scheduling UAVs and vehicles. To better adapt to the affected areas, the ground station is hosted by an emergency communication vehicle (ECV) \cite{7317860} with powerful communication and computing capabilities.

The blockchain contains a growing sequence of hash-chained blocks and each block $\mathcal{B}$ has two parts: the block body $body$ and the block header $header$.
The block body includes a set of time-stamped records (e.g., sensory data, computing results, rescue commands, and node misbehaviors) which are compressed into a Merkle tree $\mathcal{T}_1$.
To alleviate the heavy storage burden of blockchain, only the data pointers are recorded on-chain while the source data are moved to an off-chain data repository.
The block header consists of the metadata of block such as the hashes of parent and current blocks, the Merkle root of $\mathcal{T}_1$, the block height, the signature of block creator, aggregated signatures of block validators, and a script $LastProof$ for misbehavior tracing (defined in \ref{subsec:CONSENSUS}).
In RescueChain, only authorized entities can participate in the permissioned blockchain network after registration at the certification authority (CA) using their true identities. Let $\mathcal{M}= \{1,\cdots,m,\cdots,M\}$ be the set of authorized nodes in the blockchain. Three different roles of participants are considered in the blockchain as follows:
\begin{itemize}
  \item \emph{Full nodes} store the copy of all blocks and are candidates of validators.
  \item \emph{Lightweight nodes} only need to store the block headers and can receive blockchain services from nearby full nodes. They can not participate in the consensus process for ledger maintenance.
  \item \emph{Validators} are part of full nodes and serve as consensus nodes in the blockchain that are responsible for consensus management by executing consensus protocols.
\end{itemize}
In the system, each ECV acts as a full node. Each authorized UAV can opt to be a lightweight node or full node based on its computation and storage capacities.

\vspace{-1mm}
\subsection{Mobility Model}
For efficient flying trajectory modeling, the total time period $T$ is evenly divided into $N$ time slots with an interval of $\Delta_t \!=\! \frac{T}{N}$. As each time slot can be sufficiently small, UAV $j$'s instant location at $n$-th time slot can be roughly fixed.
The horizontal location of UAV $j$ at $n$-th time slot is denoted as $\mathbf{l}_j[t_n] =(x_j[t_n], y_j[t_n])$.
To avoid frequent ascending and descending and maintain continuous flight over the air, the hovering altitude of UAV $j$ is considered to be fixed at ${H}_j$ during executing a rescue task \cite{ToN9219140,8434285Jsac,7932157TVT,7888557TWC,9149835TWC}, {which varies for different tasks.} In practice, the fixed hovering altitude refers to the lowest altitude at which the UAV can avoid any terrain or building obstructions {in the task area, hence reducing the additional energy loss caused by frequent height changes}.
Let $E_j^{\min}$ be the minimum energy reserve of UAV $j$ to prolong its battery life. Then, UAV $j$'s remaining battery energy at $n$-th time slot is constrained by $E_j^{\min} \le E_j[t_n] \le C_j$, where $C_j$ is the battery energy capacity of UAV $j$.
As UAV $j$ flies at a constant height, according to \cite{7932157TVT,7888557TWC,9149835TWC}, its flying power $P_j^{\mathrm{fly}}[t_n]$ to propel its mobility and keep it aloft at $n$-th time slot is positively related with both the velocity $v_j[t_n]$ and acceleration $a_j[t_n]$, i.e.,
\begin{align}\label{eq:flyingPower}
P_j^{\mathrm{{fly}}}[t_n] = \lambda_1 (v_j[t_n])^3 + \frac{\lambda_2}{v_j[t_n]}\left(1 + \frac{(a_j[t_n])^2}{g^2} \right),
\end{align}
where $g$ is gravitational acceleration with nominal value $9.8\, \mathrm{m/s^2}$. $\lambda_1$ and $\lambda_2$ are parameters depending on UAV's weight (containing its payload), air density, wing area, UAV type (e.g., fixed-wing or rotary-wing), etc.
The flying velocity $v_j[t_n]$ of UAV $j$ is:
\begin{align}\label{eq:velocityUAV}
v_j[t_n] = \frac{\mathbf{l}_j[t_{n+1}] - \mathbf{l}_j[t_n]}{\Delta_t}\le v_j^{\max}, 1\le n < N,
\end{align}
where $v_j^{\max}$ is the maximum velocity of UAV $j$.

For ground vehicles, the fluid traffic model \cite{7094288} is adopted to capture the relationship between the traffic density $\chi$ and average vehicular velocity $\overline v_{GV}$. We have:
\begin{align}\label{eq:velocityGV}
\overline v_{GV} = \max \left\{ {{v_{GV}^{\min }},{v_{GV}^{\max }}\left( {1 - {{\chi}}/{{\chi _{\max }}}} \right)} \right\},
\end{align}
where ${\chi _{\max }}$ means the maximum traffic density. ${v_{GV}^{\max }}$ and ${v_{GV}^{\min }}$ are the maximum and minimum vehicle velocities, respectively. The number of ground vehicles entering the communication coverage of UAV $j$ at $n$-th time slot can be computed as $\phi_j[t_n] = {\chi} \overline v_{GV} \Delta_t$. Let $o_j[t_n]$ denote the ratio of ground vehicles leaving the coverage of UAV $j$ at $n$-th time slot. Based on \cite{9155397}, the number of vehicles in the communication range of UAV $j$ at $n$-th time slot is calculated as: 
\begin{align}\label{eq:vehiclenumber}
{I_j}[t_n] \!=\! \begin{cases}
{\phi _j}[t_1] (1 - {o_j}[t_1]), &n=1;\\
\left({\phi _j}[t_n] + {I_j}[t_{n-1}]\right) \left(1 - {o_j}[t_n]\right),&1 \!<\! n \!\le\! N.
\end{cases}
\end{align}

\subsection{Vehicular Fog Computing Model}
The limited communication and battery resources of UAVs may be unaffordable to execute compute-intensive and delay-critical rescue tasks. Consequently, the executing time for rescue missions may be prolonged and the recharging interval of UAVs may be shortened.
Under the VFC paradigm, the UAV's heavy data computation and storage missions can be efficiently offloaded to ground vehicles in its communication coverage by collaboratively contributing their idle resources. Let $\mathcal{I}_j[t_n] = \{1,\cdots,i,\cdots,I_j[t_n]\}$ be the set of cooperative vehicles in UAV $j$'s coverage at $n$-th time slot.
Each task of UAV $j$ to be offloaded to a vehicle $i \in \mathcal{I}_j[t_n]$ can be denoted as a 4-tuple, i.e.,
\begin{align}\label{eq:task}
\Gamma_{j,k} = \left\langle D_{j,k}, \theta_{i,j,k}, T_{j,k}^{\max}, \alpha_{j,k} \right\rangle, 1\le k \le K_j,
\end{align}
where $D_{j,k}$ (in bits) denotes the data size of task. $\theta_{i,j,k}$ (in CPU cycles/bit) indicates the required CPU cycles to process one bit. $T_{j,k}^{\max}$ (in seconds) means the time-to-live (TTL) of task. $\alpha_{j,k} \in [0,1]$ is the urgency degree for mission processing, and a higher $\alpha_{j,k}$ implies a higher urgency degree. $K_j$ is the total number of UAV $j$'s tasks to be offloaded.

In VFC, when all $D_{j,k}$ bits of task data are transmitted to a ground vehicle $i \!\in\! \mathcal{I}_j[t_n]$ through A2G communication, the data computation process is executed immediately on the vehicle, and then $\mu_{j,k} D_{j,k}$ bits of the processed results are sent back to the UAV $j$ through ground-to-aerial (G2A) communication. Here, $\mu_{j,k} \!\in\! (0,1)$ denotes the output/input ratio.
Let $x_{i,j,k}$ be the amount of computing resource (AoCR) contributed by vehicle $i$ in performing mission $\Gamma_{j,k}$, which is measured by CPU cycles per second.
According to \cite{8956055}, the execution time of mission $\Gamma_{j,k}$ in vehicle $i$ can be calculated as
$t_{i,j,k}^{\mathrm{vfc}} =\frac{ \theta_{i,j,k}D_{j,k}}{x_{i,j,k}}$.
Besides, the energy consumption of vehicle $i$ in performing mission $\Gamma_{j,k}$ is:
\begin{align}\label{eq:executionenergy}
E_{i,j,k}^{\mathrm{vfc}} = \kappa_i (x_{i,j,k})^3 \cdot t_{i,j,k}^{\mathrm{vfc}} = \kappa_i \theta_{i,j,k}D_{j,k} (x_{i,j,k})^2,
\end{align}
where $\kappa_i$ is the effective switched capacitance \cite{8908666}.

\vspace{-0.2cm}
\subsection{Communication Model}
For A2G/G2A communications, the line-of-sight (LoS) link is presumed to be dominant \cite{8956055}, and the channel gain between UAV $j$ and ground vehicle $i$ can be modeled by the free-space path loss model, i.e., $\varphi _{i,j}[t_n] = {\varphi _0}\left(d_{i,j}[t_n]\right)^{ - \mu}$, where ${\varphi _0}$ is the channel gain at the reference distance $d_0=1$, $\mu>1$ means the path loss exponent, and $d_{i,j}[t_n]$ represents the Euclidean distance between UAV $j$ and ground vehicle $i$ at $n$-th time slot.
The frequency division multiple access (FDMA) protocol is applied for bandwidth sharing among ground vehicles in transmitting computation outcomes to the UAV \cite{8908666}.
Then, the available G2A uplink transmission rate from vehicle $i$ to UAV $j$ is obtained as:
\begin{align}\label{eq:uplinkrate}
{\gamma _{i,j}^{\mathrm{G2A}}}[t_n] = B_{i}^{\mathrm{UL}}{\log _2}\bigg( {1 + \frac{{P_i^{\mathrm{TX}}{\varphi _0}\left(d_{i,j}[t_n]\right)^{ - \mu} }}{B_{i}^{\mathrm{UL}} \sigma_0^2}} \bigg),
\end{align}
where $B_{i}^{\mathrm{UL}}$ is the assigned uplink bandwidth to vehicle $i$, $P_i^{\mathrm{TX}}$ means vehicle $i$'s transmit power, and $\sigma_0^2$ is the power of additive white Gaussian noise.
Moreover, the A2G downlink transmission rate from UAV $j$ to vehicle $i$ is denoted as:
\begin{align}\label{eq:downlinkrate}
{\gamma _{i,j}^{\mathrm{A2G}}}[t_n] = B_{j}^{\mathrm{DL}}{\log _2}\bigg( {1 + \frac{{P_j^{\mathrm{TX}}{\varphi _0}\left(d_{i,j}[t_n]\right)^{ - \mu} }}{B_{j}^{\mathrm{DL}} \sigma_0^2}} \bigg),
\end{align}
where $B_{j}^{\mathrm{DL}}$ is UAV $j$'s downlink bandwidth, and $P_j^{\mathrm{TX}}$ is UAV $j$'s transmit power.
\vspace{-0.2cm}
\subsection{Adversary Model}\label{subsec:threatmodel}
Three kinds of adversaries considered during information sharing in UDRNs are listed as follows.
\begin{itemize}
  \item \emph{{Malicious Insider UAVs}.} Due to the open-access features of UAV networking to maximize the life-saving potential under disasters, any entities including malicious ones can connect to UAVs freely. Consequently, UAVs may be compromised and controlled by them to tamper with the sensory data, disseminate falsified information, and produce fake blocks.
  \item \emph{{Threats to Data Dependability}.} Owing to the unreliable wireless environment, the disseminated data (e.g., sensory data, processed results, and scheduling commands) may be forged, replaced, or deleted during multi-hop data transmissions{\cite{9687566}}. Moreover, it is difficult to transparently audit the delivered information to enforce accountability when disputes occur.
  \item \emph{{Threats to Misbehavior Traceability}.} Traditional centralized misbehavior tracing mechanisms usually lack transparency and auditability and may suffer from the single point of failure (SPoF) and DDoS attack, causing difficulty in trustworthy, auditable, and immutable misbehavior tracing and punishment enforcement.
\end{itemize}

\section{Design of RescueChain}\label{sec:SOLUTION1}
In this section, we present the detailed design of RescueChain, which is a blockchain-based system for secure and efficient information sharing in UDRNs.
\vspace{-0.2cm}
\subsection{Overview of RescueChain}\label{subsec:OVERVIEW}
\begin{figure}[t!]\setlength{\abovecaptionskip}{-0.05cm}
\centering
\setlength{\abovecaptionskip}{-0.02cm}
  \includegraphics[width=9.cm]{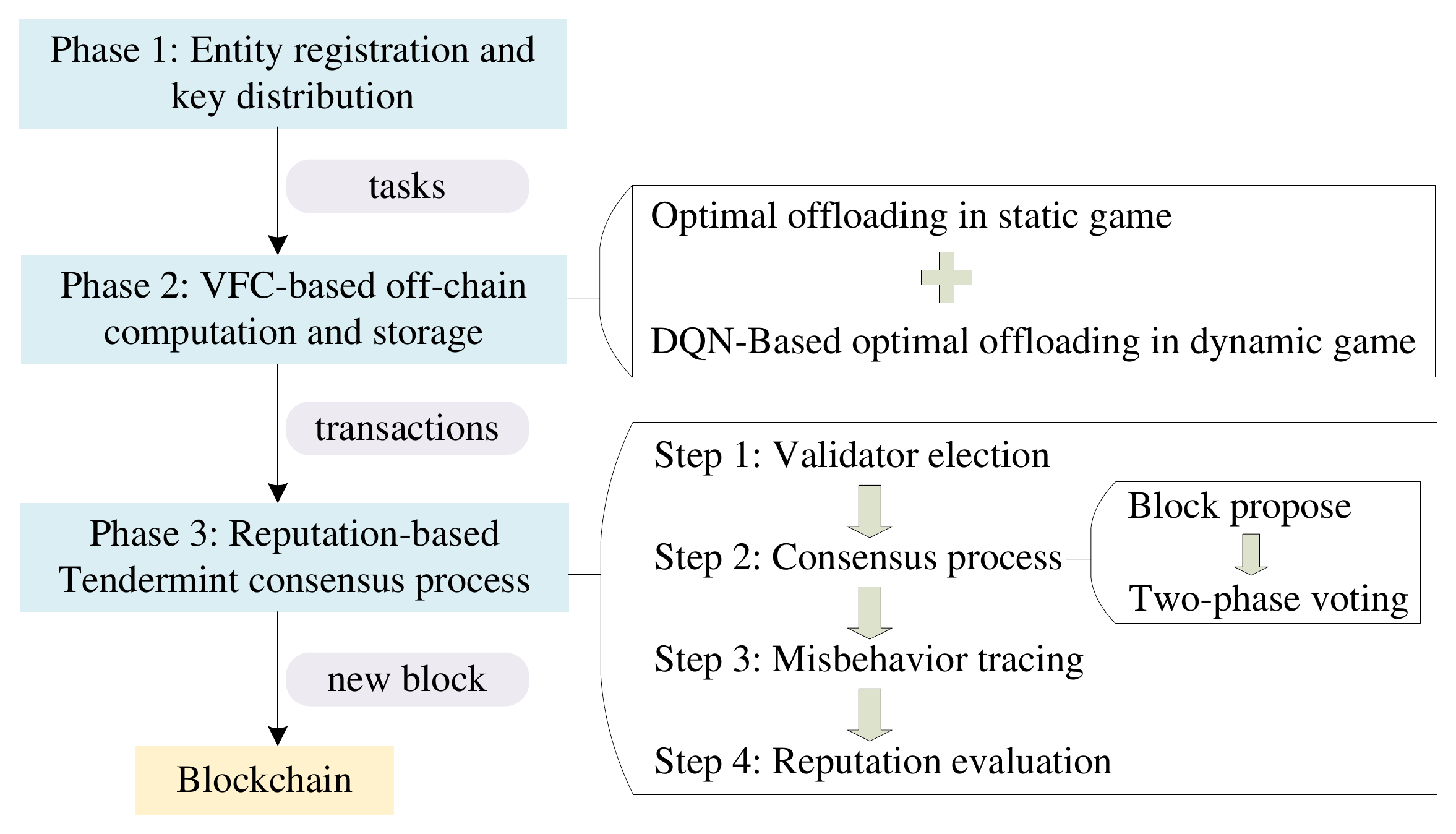}
  \caption{{Workflow of the proposed Rescuechain.}}\label{fig:Workflow}\vspace{-0.3cm}
\end{figure}

In UDRNs, owing to the sparse network environment and possible poor connections, network partitions may be frequent, resulting in the risks of blockchain forking.
Besides, the limited resource of UAVs in terms of battery, computation, and storage restrict the adoption of conventional resource-hungry blockchains. Accordingly, our RescueChain should be \emph{low-cost}, \emph{infrastructure-free}, and \emph{robust} in UDRNs. To achieve the three goals, we make two improvements: one is the VFC-based off-chain mechanism to offload the heavy data computation and storage tasks from UAVs; the other is the reputation-based Tendermint consensus protocol to robustly reach consensus in the weakly connected environment with enhanced safety.
In particular, {as shown in Fig.~\ref{fig:Workflow},} the following three phases need to be undertaken:

\begin{itemize}
  \item \emph{Entity registration and key distribution.} In this phase, after registration at CA, each authorized entity obtains its key pair and opts its role in the blockchain system.
  \item \emph{VFC-based off-chain storage and computation.} This phase performs off-chain data computation and storage for UAVs under the VFC paradigm.
  \item \emph{Reputation-based Tendermint consensus process.} In this phase, each entity runs the reputation-based Tendermint protocol to reach consensus on the new transactions to be added to the blockchain via validator election, two-phase voting with locking, and reputation assessment.
\end{itemize}

\vspace{-0.3cm}
\subsection{Entity Registration and Key Distribution}\label{subsec:REGISTER}
The system parameters are selected by CA based on the Boneh-Lynn-Shacham (BLS) short signature scheme \cite{10.1007/BLS}. A multiplicative bilinear map $e: \mathbb{G}_1 \times \mathbb{G}_2 \rightarrow \mathbb{G}_3$ is chosen by CA, where $g_1$ and $g_2$ are generators of groups $\mathbb{G}_1$ and $\mathbb{G}_2$, and $q$ is the prime order of $\mathbb{G}_k (k=1,2,3)$. Two hash functions $\mathbf{H}_0:\{0,1\}^*\rightarrow \mathbb{G}_2$ and $\mathbf{H}_1:\{0,1\}^*\rightarrow \mathbb{Z}_q$ are chosen by CA.
Then, CA broadcasts the system parameters $parm$ to the network:
\begin{align}\label{4-para}
parm = (e,\mathbb{G}_1,\mathbb{G}_2,\mathbb{G}_3,q,g_1, g_2,\mathbf{H}_0,\mathbf{H}_1 ).
\end{align}
Each authorized node $m \in \mathcal{M}$ registers itself at CA using the real identity $RID_m$ and receives a group of $L$ private/public key pairs $\{{sk}_m^l,{pk}_m^l\}_{l=1}^L$, wallet addresses, and certificates in the blockchain from CA. Here, ${sk}_m^l \xleftarrow{S} \mathbb{Z}_q$, ${pk}_m^l = g_2^{{sk}_m^l}\in \mathbb{G}_2$, and $\xleftarrow{S}$ indicates randomly sampling.

\subsection{VFC-Based Off-Chain Storage and Computation}\label{subsec:OFFCHAIN}
Under VFC, UAVs' compute-intensive missions (e.g., human detection and multi-object tracking) on sensory data can be offloaded to ground vehicles to save their constrained battery energy. To alleviate the overhead of blockchain and improve system scalability, the VFC-based off-chain approach is adopted by moving both sensed and processed data to off-chain repositories while retaining a data pointer of source data on the blockchain. The set of data to be disseminated in UDRNs is denoted as $\mathcal{D}$. Each data $d \in \mathcal{D}$ is composed of two parts, i.e., the raw sensory data $d_{\mathrm{raw}}$ and the computation outcome $d_{\mathrm{out}}$.
In RescueChain, the InterPlanetary File System (IPFS) \cite{9045940} serves as the distributed off-chain data store, where each data $d$ is stored in the format of distributed files in IPFS and is uniquely addressed by its hash pointer $\mathbf{H}_0(d) = \{\mathbf{H}_0(d_{\mathrm{raw}}),\mathbf{H}_0(d_{\mathrm{out}})\}$.
The ground vehicles serve as the distributed storage nodes in IPFS.
After VFC-based computation offloading process for task $\Gamma_{j,k}$, UAV $j$ sends an off-chain storage transaction $tx_{\mathrm{data}}$ to the network as:
\begin{align}\label{4-off-chain}
  &tx_{\mathrm{data}} = \nonumber \\
  &\resizebox{0.894\hsize}{!}{$ \bigm\langle {pk}_j^l||\{{pk}_i^l\}_{i=1}^{I_j} ||\mathbf{H}_0(d)||desc_d||tStamp||\sigma_j ||\sigma_k||cert_{DS} \bigm\rangle, $}
\end{align}
where ${pk}_j^l$ is the public key of UAV $j$. $\{{pk}_i^l\}_{i=1}^{I_j}$ is the group of public keys of vehicles involved in task $\Gamma_{j,k}$. $desc_d$ is the description of data $d$. $tStamp$ is the timestamp of $tx_{\mathrm{data}}$ generation. $\sigma_j = \mathbf{H}_0(tx_{\mathrm{data}})^{{sk}_j^l}$ means the signature of UAV $j$. $\sigma_k = \prod_{i=1}^{I_j}\bar{\sigma}_i$ is the BLS multi-signature of ${I_j}$ vehicles, where $\bar{\sigma}_i = \mathbf{H}_0(tx_{\mathrm{data}})^{\omega_i\cdot {sk}_i^l}$ and $\omega_i = \mathbf{H}_1({pk}_i^l,\{{pk}_1^l,\cdots,{pk}_{I_j}^l\})$. $cert_{DS}$ is the unique certificate issued by the IPFS data store.

\subsection{Reputation-Based Tendermint Consensus Protocol}\label{subsec:CONSENSUS}
In RescueChain, the Tendermint protocol \cite{Kwon2014Tendermint} is adopted to reach consensus among distrustful UAVs due to its high energy efficiency and partial partition tolerance than its alternatives.
To be adaptive to the sparse network environment in UDRNs, the multi-signature method \cite{10.1007/BLS} is applied in the consensus phase for bandwidth saving.
To further enhance the safety of consensus and speed up the consensus process, our RescueChain incorporates reputation assessment into the Tendermint for validator election and block bookkeeping. In the proposed reputation-based Tendermint protocol, the following steps are included:

\textbf{Step 1: Validator election.}
The consensus process is carried out by an elected validator committee, denoted as $\mathcal{Z}= \{1,\cdots,z,\cdots,Z\}$. Let $\mathcal{U}= \{1,\cdots,u,\cdots,U\}$ be the set of full nodes. Every full node $u\in\mathcal{U}$ can vote for a delegate and its voting weight is determined by its stake, namely, the reputation value ${\overline{\Im}_u}$.
The top $Z$ delegates with the highest votes are selected to form the validator committee $\mathcal{Z}$, which can be further classified into two kinds: level-1 validators and level-2 validators. The top $\Psi$ validators with the highest votes serve as level-1 validators which can propose and validate new blocks, while the remaining $Z-\Psi$ validators are level-2 validators that can only perform block verification.

\textbf{Step 2: Consensus process.}
The consensus process is composed of three phases: \emph{propose}, \emph{prevote}, and \emph{precommit}.
To eventually reach consensus in weakly connected UDRNs, the network is assumed to be partially synchronous, and the partial synchrony model \cite{PartialSynchrony} is applied in protocol design. An upper bound $\Delta$ of transmission delay and an unknown global stabilization time (GST) are defined in UDRN. In the partial synchrony model, there exist either a known upper bound $\Delta$ after the unknown GST or an unknown upper bound $\Delta$ on the transmission latency of messages.
It means that the transmission between two correct entities in UDRN will eventually arrive within 1) the unknown timeout $\Delta$, or 2) the known timeout $\Delta$ beginning with an unknown GST.
At each round of consensus process for a certain height, a small fixed increment of the timeout $\Delta$ is added than its previous round \cite{Buchman2016Tendermint}. Besides, the internal clocks of non-Byzantine validators are assumed to be sufficiently accurate during a short time period until reaching consensus on the next block. In the following, we elaborate the detailed process to reach consensus in our RescueChain.

\emph{1) Propose phase.} A leader $\hbar$ is designated from $\Psi$ level-1 validators in a round-robin fashion. At its round $r$, the leader $\hbar$ collects a batch of recent transactions from its local memory pool, compresses them into a Merkle tree $\mathcal{T}_1$, and packages them into a block $\mathcal{B}$ which includes a hash linked to the previous block. Then, a signed proposal is gossiped to the network, i.e.,
\begin{align}\label{eq:proposalMsg}
{proposal} = \left\langle {{\mathrm{Proposal}}||h|| r|| \mathcal{B}|| \Upsilon_{\hbar} ||\sigma_{\hbar} }\right\rangle,
\end{align}
where $\Upsilon_{\hbar}$ is the proof-of-lock (PoL) indicating the locked block of the leader ${\hbar}$, and $\sigma_{\hbar}$ is its signature. A PoL for block $\mathcal{B}$ (or $nil$) means that a leader receives at least $\frac{2}{3}Z$ prevotes for block $\mathcal{B}$ (or $nil$) at round $r$ for height $h$ \cite{Buchman2016Tendermint}. Note that the block $\mathcal{B}$ can be quite large, and can be a network bottleneck in disseminating it to other validators.
To alleviate the communication burden of leader ${\hbar}$ in disseminating block $\mathcal{B}$, we design a block partition method for efficient proposal transmission. Specifically, the raw block is serialized and split into multiple chunks with appropriate size, and all the chunks are hashed into a Merkle tree $\mathcal{T}_2$. A Merkle root $root_{\mathcal{B}}$ together with the signature of leader ${\hbar}$ are included in the proposal. Once a validator receives all the chunks, it performs the block deserialization operation and validates the correctness of the received proposal by checking the Merkle root $root_{\mathcal{B}}$.

\emph{2) Two-phase voting with locking.} In two-phase voting, each validator sends its prevote result of the proposal to the network in the first phase, and tells its precommit result of other validators' claims regarding the proposal in the second phase.
Intuitively, it can be ensured that the result of the first phase has been witnessed by enough validators in the second phase to offer Byzantine tolerance.
Besides, the following two rules are defined in the locking mechanism to ensure blockchain safety:
\begin{itemize}
  \item \emph{Locking rule.} A validator $z$ is locked on ${\mathcal{B}}$ (or $nil$) if it receives over $\frac{2}{3}Z$ prevotes for block $\mathcal{B}$ (or $nil$) at round $r$ for height $h$. Then, at round $r' > r$ for height $h$, it should prevote for the locked block, and propose it if it is the leader. This can prevent validators from prevoting different blocks in different rounds, thereby protecting safety as conflicting blocks may be committed at the same height.
  \item \emph{Unlocking rule.} A validator can only unlock the block ${\mathcal{B}}$ which is locked at round $r$ if there exist a PoL for another block ${\mathcal{B}}'$ (or $nil$) at a higher round $r'$, where $r' > r$. This allows validators to be unlocked and precommit another block that the rest of the network intend to commit, thereby protecting liveness.
\end{itemize}

If the validator $z \in \mathcal{Z}$ is locked on a block proposal at a previous round, it broadcasts a prevote for the locked block with its signature $\sigma_{z}$. Otherwise, if a valid proposal is received for the current round after verification, validator $z$ broadcasts a prevote for that block with its signature.
If no proposal or a invalid one is received within the timeout $proposeTimeout$, validator $z$ sends a special prevote for $nil$ instead. The form of prevote message is:
\begin{align}\label{eq:prevoteMsg}
{prevote} = \left\langle {{\mathrm{prevote}}||h|| r|| \mathbf{H}_0(block)|| \sigma_{z}}\right\rangle,
\end{align}
where $block$ is the prevoted block or $nil$. $proposeTimeout$ is the maximum time that a validator stays in the propose phase, which is initialized as $\Delta_{propose}$ at the beginning of a height $h$ and is incremented every time $proposeTimeout$ expires.

If there exists a PoL for block $\mathcal{B}$, the validator $z$ locks on block $\mathcal{B}$, releases the prior locked block, and broadcasts a precommit for $\mathcal{B}$. If there exists a PoL for $nil$, the validator $z$ unlocks, and it broadcasts a precommit for $nil$. Otherwise, the validator $z$ precommits $nil$. The explicit form of precommit message is
\begin{align}\label{eq:precommitMsg}
{precommit} = \left\langle {{\mathrm{precommit}}||h|| r|| \mathbf{H}_0(block)|| \sigma_{z}}\right\rangle.
\end{align}
The same prevote or precommit messages signed by distinct validators can be aggregated by BLS multi-signatures for compressed signature size.
If validator $z$ receives over two-thirds of precommits for a specific block, it simply sets its commit time $commitTime$ to the current time and moves on to round $0$ at the next height $h+1$. Then, the specific block is eventually committed. Otherwise, it moves to the next round $r+1$ at current height $h$. Then, a new leader is designated and the above process is repeated.
Before entering the next height $h+1$, validators wait until some fixed time duration past $commitTime$ to include more commits from validators with slower network connections.

\textbf{Step 3: Misbehavior tracing.} Malicious validators may conduct various kinds of misbehaviors to crash the consensus process and fork the blockchain.
A malicious leader may conduct \emph{conflicting proposals (cp)} misbehavior by disseminating conflicting block proposals to different validators within a round, and malicious validators may perform \emph{conflicting votes (cv)} misbehavior by giving prevotes or precommits to conflicting proposals.
A malicious leader can conduct \emph{wrong block creation (wbc)} misbehavior by producing invalid block or true block at false round and height, and \emph{non-block creation (nbc)} misbehavior if the block proposal is not generated on time or it is offline or partitioned.
The above misbehaviors (i.e., $cp$, $cv$, $wbc$, and $nbc$) can be identified via their signatures attached on the proposals or votes, thereby the corresponding misbehaving entities can be detected.
Moreover, malicious validators can perform \emph{violation of locking (vol)} misbehaviors by violating the locking and unlocking rules to make the network commit on two different blocks at a height.
In this case, if a prevote or precommit resulted from a $vol$ misbehavior affects the final commit, at least one honest validator must have been received it. As honest validators should broadcast all their received votes at each round of a height, the evidences of violations of the locking rule can be collected by stitching all votes and matching every prevote with the most recent precommit of the same validator. The violations of unlocking rule can be detected similarly via matching every precommit with the PoL that justifies it.

For efficient and reliable misbehavior forensics and tracing, a script denoted by $LastProof$ is included in the next block at height $h+1$, which contains the set of signed precommits that make the prior block at height $h$ committed (i.e., $LastCommit$), together with all votes justifying the precommits at all rounds for height $h$ (i.e., $LastVotes$). As such, the unjustified votes and the misbehaving entities can be identified and traced.
For alleviated blockchain burden, all the votes in $LastVotes$ are moved to an off-chain data store instead, and only a Merkle tree $\mathcal{T}_3$ constructed from $LastVotes$ are included in the $LastProof$ with a root $root_V$.
Furthermore, to accelerate misbehavior detection, nodes can form informer groups to report misbehaviors by collecting the cryptography trails as evidence and generating a report transaction as:
\begin{align}\label{4-reportmisb}\resizebox{0.894\hsize}{!}{$
  tx_{\mathrm{rep}} = \bigm\langle {pk}_{u'}^l||\{{pk}_u^l\}_{u=1}^{N_{rep}} ||Evidence||fee|| tStamp||\sigma_{rep} \bigm\rangle, $}
\end{align}
where ${pk}_{u'}^l$ is the public key of accused node $u'$.
$\{{pk}_u^l\}_{u=1}^{N_{rep}}$ and $\sigma_{rep}$ are the public keys and BLS multi-signature of the ${N_{rep}}$ informers, respectively. $fee$ is the fixed report fee to prevent DDoS attack and is evenly divided among them. $Evidence$ is the collected evidence such as the digital signatures on votes or proposals at different rounds of a height.

\textbf{Step 4: Reputation evaluation.}
The reputation of each full node is calculated based on two aspects: behavior effect and time fading effect.
Commonly, node's reputation can be gradually improved by benign behaviors and significantly decreased by misbehaviors; meanwhile, latest behaviors are of more significance than previous ones in determining current reputation.
The following benign behaviors are considered in blockchain: the \emph{successful block creation (sbc)} behavior of the leader and the \emph{successful block verification (sbv)} behavior of validators.
Both the $sbc$ and $sbv$ behaviors can be detected by the signatures on block proposal and precommits included in the $LastCommit$.
Let $\Delta_b$ be the reputation reward or penalty for behavior $b$. A binary variable $\beta_b = \{-1,1\}$ is defined, where $\beta_b = 1$ if $b$ is a benign behavior; and $\beta_b \!=\! -1$ if $b$ is a misbehavior.
Besides, if $tx_{\mathrm{rep}}$ is verified as true and recorded in blockchain, all members within the informer group are rewarded with a reputation increase $\Delta_{rep}$, as well as a reputation decrease $\Delta_{acc}$ for the accused entity. Otherwise, the expended report fee is in vain. Here, the evidence collection and misbehavior forensics are out of scope for this paper, and readers can refer to existing works \cite{ToN8264787,DF7879878}.
At each time slot, each full node $u$'s reputation can be computed as:
\begin{align}\label{4-credit}
\Im_u^{n} =
\begin{cases}
\Im_u^{\mathrm{ini}},
& n=1;\\
\sum\nolimits_{b = 1}^{N_u^n}{ \beta_b \Delta_b } + e^{-\eta} \Im_u^{n-1},
& 1<n \le N,
\end{cases}
\end{align}
where $\Im_u^{\mathrm{ini}}$ is the initial reputation of full node $u$, ${N_u^n}$ is the number of recorded behaviors of full node $u$ at $n$-th time slot, and $\eta>0$ is the exponential decay factor. The normalized reputation of each full node can be formulated as ${{\overline{\Im}_u^{n}}} = \left( 1+e^{-\Im_u^{n}} \right)^{-1}$.
The reputation values of full nodes are updated after each round of consensus according to the time-stamped behavior records stored in the blockchain.

\subsection{Analysis of RescueChain}\label{subsec:SECURITYANALYSIS}
Let $P_b$ denote the ratio of Byzantine validators in RescueChain, and $P_p$ be the ratio of validators that are offline or partitioned in UDRNs.
According to \cite{Buchman2016Tendermint,Kwon2014Tendermint}, under partial synchrony assumption of the network, our RescueChain system is resilient up to $N_{fault} = \lfloor \left( Z-1 \right) /3 \rfloor$ Byzantine validators. Next, we analyze the finality, network availability, and communication complexity of RescueChain in the following Theorems 1--3, respectively.

\emph{\textbf{ Theorem 1 (Finality)}: }
With assumptions of partial synchrony and $P_b \le 1/3$, by taking multiple rounds at a given height to commit a block via two-phase voting with locking, our RescueChain can reach deterministic finality, where a deadlock that different blocks are locked by part of honest validators at different rounds can be avoided.
\begin{IEEEproof}
We prove the theorem by contradiction. Suppose that RescueChain does not satisfy finality, where more than one block are committed at the same height.
Without loss of generality, we consider two committed blocks $\mathcal{B}$ and $\mathcal{B}'$ of height $h$ at rounds $r$ and $r'$, respectively.
There exist the following two cases: $r' = r$ and $r' \ne r$. In the first case that $r' = r$, both blocks $\mathcal{B}$ and $\mathcal{B}'$ receive over two thirds of precommits at round $r$. Obviously, at least $1/3$ Byzantine validators has precommitted for both blocks at round $r$.
In the other case that $r' \ne r$, we assume $r' > r$ without loss of generality. As over two thirds of validators precommit for $\mathcal{B}$ at round $r$, they get locked on $\mathcal{B}$ and must prevote for $\mathcal{B}$.
To precommit for block $\mathcal{B}'$, it requires over two-thirds to prevote for $\mathcal{B}'$, which means that at least $1/3$ Byzantine validators violate the locking and unlocking rules.
\end{IEEEproof}

\emph{\textbf{ Theorem 2 (Availability)}: }
In face of network asynchrony, our system will never halt if $1 - P_b - P_p\cdot \left(1-P_{pb}\right) > 2/3$, where $P_{pb}$ is the ratio of offline or partitioned Byzantine validators to the total number of offline or partitioned validators.
\begin{IEEEproof}
Let $P_n^r$ denote the ratio of remaining non-Byzantine validators that are not offline or partitioned in UDRNs. We have $P_n^r = 1 - P_b - P_p\cdot \left(1-P_{pb}\right)$. When $P_n^r> 2/3$, as analyzed in Theorem~1, our system will never halt and the unambiguous consensus can be reached.
\end{IEEEproof}

\emph{\textbf{Remark:}}
If over a third of validators get offline or partitioned (i.e., $P_p > 1/3$), the network will halt since no block can receive two thirds or more votes from non-Byzantine validators (i.e., $P_n^r< 2/3$). The unambiguous consensus can be restored until the network connection recovers.

\emph{\textbf{ Theorem 3 (Complexity)}: }
Compared with the $\mathcal{O}(Z^3)$ view-change overhead in PBFT, the total communication complexity is reduced to $\mathcal{O}(Z^2)$ in our RescueChain.
\begin{IEEEproof}
In the propose phase, the new block proposal needs to be delivered to other $Z\!-\!1$ validators, and the communication complexity is $\mathcal{O}(Z)$. In the two-phase voting phase, each validator needs to collect the voting messages from others to make progress, and the communication complexity is $\mathcal{O}(Z^2)$. Therefore, the total communication complexity is $\mathcal{O}(Z^2)$.
\end{IEEEproof}

\emph{\textbf{Remark:}}
Based on the proposed block partition method in the propose phase and the BLS signature aggregation in two-phase voting, the delivery latency of block proposal and the signature size in voting can be reduced, thereby alleviating the communication burden in consensus process.

Next, we give the security analysis of our RescueChain.
Our paper mainly focuses on the blockchain-based solution to defend against attacks defined in Sect. \ref{subsec:threatmodel}. The authentication and access control, which are also essential to ensure network security in UDRNs, are out of scope for this paper, and readers can refer to existing cryptographic approaches \cite{ToN8630993,ToN8635539}.
\begin{itemize}
  \item Thanks to the decentralized ledgers and special data structure of blockchain (i.e., transactions are time-stamped, signed, and recorded in the Merkle tree structure in hash-chained blocks), the metadata of shared data can be immutably recorded on the blockchain to prevent data tampering attacks and guarantee the \emph{integrity and dependability of delivered information}; meanwhile, the recorded node misbehaviors on blockchain ledgers can be transparently traced back to the source to ensure the \emph{traceability of misbehaviors}. Besides, the defense against \emph{compromised insider UAVs} is analyzed in the simulation section.
  \item Since only registered nodes are authorized for network access, the \emph{air traffic scheduling} of UAVs can be facilitated while \emph{preventing identity thefts}.
      By dynamically altering the public keys in transactions via existing pseudonymity mechanisms in \cite{ToN8884665}, the true identities can be hidden and remain unlinkable to preserve nodes' \emph{identity privacy}.
\end{itemize}

\section{Optimal Offloading in Static and Dynamic Games}\label{sec:SOLUTION2}
In this section, the payoff functions and optimization problems of UAVs and ground vehicles in VFC-based offloading are first formulated. Then, we analyze the optimal offloading strategies of vehicles and the optimal payment strategies of UAVs under a static Stackelberg game model with one interaction and a dynamic Stackelberg game with repeated interactions, respectively.

\subsection{Optimization Problem}\label{subsec:problemformulation}
\underline{Payoff function of UAV.}
During VFC-based offloading process, to stimulate ground vehicles' participation and high amount of resource sharing, each UAV $j$ needs to determine the payment vector ${\bf{y}}_{j} = \left\{{y_{i,j,k}}\right\}_{k=1}^{K_j}$ for each task $k$ to compensate the cost of ground vehicles in contributing computation resources. Let ${\bf{x}}_{j} = \left\{{x_{i,j,k}}\right\}_{k=1}^{K_j}$ denote the AoCR vector of ground vehicles that participate in the offloading process for UAV $j$.
The payoff function of UAV $j$ is defined as the difference between the satisfaction and its cost, i.e.,
\begin{align}\label{eq:payoffUAV}\resizebox{0.894\hsize}{!}{$
{\pi _j}({\bf{x}}_{j},{\bf{y}}_{j}) \!=\! \sum\limits_{k = 1}^{{K_j}}{\sum\limits_{i = 1}^{{I_j}} {\beta _{i,j,k}\left[S_{j,k}\left(x_{i,j,k} \right) \!-\! C_{j,k}\left(x_{i,j,k},y_{i,j,k} \right)  \right] }}, $}
\end{align}
where $\beta _{i,j,k}\!=\!\{0,1\}$ is a binary variable which equals to zero if vehicle $i$ does not participate in the offloading of task $\Gamma_{j,k}$, otherwise it equals to one. $S_{j,k}\left(x_{i,j,k} \right)$ is the satisfaction function of UAV $j$ with the obtained AoCR ${x_{i,j,k}}$ of each task $\Gamma_{j,k}$. $C_{j,k}\left(x_{i,j,k},y_{i,j,k} \right)$ is the cost function of UAV $j$ during offloading. Here, the logarithmic function, which is widely employed in resource allocation mechanisms \cite{7275166}, is used to model the satisfaction. We have
\begin{align}\label{eq:satisfaction}
S_{j,k}\left(x_{i,j,k} \right) = {\rho _{j}}{\alpha _{j,k}}\log \left( {1 + {x_{i,j,k}}} \right),
\end{align}
where ${\rho _j}$ means the satisfaction parameter of UAV $j$, and ${\alpha _{j,k}}$ is the urgency degree of task $\Gamma_{j,k}$ defined in Eq. (\ref{eq:task}). The cost function $C_{j,k}\left(x_{i,j,k},y_{i,j,k} \right)$ is composed of two parts: the payment for shared computation resource (i.e., ${\lambda_p {x_{i,j,k}}{y_{i,j,k}}}$) and the time delay to obtain the processing results (i.e., $\Phi_{i,j,k}$). We have
\begin{align}\label{eq:costUAV}\resizebox{0.894\hsize}{!}{$
C_{j,k}\left(x_{i,j,k},y_{i,j,k} \right) =  {{\varpi_p}\lambda_p{y_{i,j,k}}{x_{i,j,k}} + \left({1 - {\varpi_p}}\right) \Phi_{i,j,k} }, $}
\end{align}
where ${\varpi_p}$ means the weight factor, and $\lambda_p$ is the price adjustment parameter. Next, we analyze the total delay and energy consumption in task offloading.

\emph{1) Delay analysis.}
The total offloading latency $\Phi_{i,j,k}$ for every task $\Gamma_{j,k}$ includes three parts: the A2G transmission time $t_{i,j,k}^{\mathrm{A2G}}$, the task execution time $t_{i,j,k}^{\mathrm{vfc}}$, and the G2A transmission time $t_{i,j,k}^{\mathrm{G2A}}$.
The A2G transmission time is denoted as $t_{i,j,k}^{\mathrm{A2G}} = \frac{{{D_{j,k}}}}{{\gamma _{i,j}^{\mathrm{A2G}}}}$.
As UAVs and ground vehicles are highly mobile, vehicle $i$ may leave the communication range of UAV $j$ and drive into the coverage of another UAV $j'$, $j' \ne j$. Without loss of generality, we denote $t_k$ as the time slot that UAV $j$ begins to offload mission $\Gamma_{j,k}$. Two cases of G2A transmission latency are considered as below: 
\begin{align}\label{eq:timeG2A}
t_{i,j,k}^{\mathrm{G2A}} = \begin{cases}
\frac{{\mu_{j,k}{D_{j,k}}}}{{\gamma _{i,j}^{\mathrm{G2A}}}},&\resizebox{0.4\hsize}{!}{${\rm{if}}~i \in {\mathcal{I}_j}\left[ {t_k} + t_{i,j,k}^{\mathrm{A2G}} + t_{i,j,k}^{\mathrm{vfc}} \right]\!;\!$}\\[0.18cm]
\frac{{\mu_{j,k}{D_{j,k}}}}{{\gamma _{i,{j'}}^{\mathrm{G2A}}}}+\frac{{\mu_{j,k}{D_{j,k}}}}{{\gamma _{{j'},j}^{\mathrm{A2A}}}},&{\rm{otherwise}},
\end{cases}
\end{align}
where ${\gamma _{{j'},j}^{\mathrm{A2A}}}$ is the A2A transmission rate between UAV $j$ and UAV $j'$. Hence, we have $\Phi_{i,j,k} = t_{i,j,k}^{\mathrm{vfc}} + t_{i,j,k}^{\mathrm{A2G}} + t_{i,j,k}^{\mathrm{G2A}}$.

\emph{2) Energy consumption analysis.}
The energy consumption of UAV $j$ during task offloading includes two parts: the A2G transmission energy and the flying energy.
UAV $j$'s energy consumption during A2G transmission is denoted as $E_{i,j,k}^{\mathrm{A2G}} = P_j^{\mathrm{TX}}\, t_{i,j,k}^{\mathrm{A2G}}$.
The flying energy of UAV $j$ can be attained as
$E_{i,j,k}^{\mathrm{fly}} = P_j^{\mathrm{fly}}\Phi_{i,j,k}$.

The target of each UAV $j\in \mathcal{J}$ is to maximize its payoff function by deciding the optimal payment strategy ${{\bf{y}}_{j}}$, whereby its optimization problem $P1$ is formulated as:
\begin{align}\label{eq:payoffUAV2}
\mathop {\max }\limits_{{\bf{y}}_{j}}\,\pi _j\left(\mathbf{x}_j, \mathbf{y}_j\right) &=\sum\limits_{k=1}^{K_j}{\sum\limits_{i=1}^{I_j}{\pi _j\left({x}_{i,j,k}, {y}_{i,j,k}\right)}}  \nonumber \\
&= \sum\limits_{k=1}^{K_j}{\sum\limits_{i=1}^{I_j}{\beta _{i,j,k}\left\{ \rho _j\alpha _{j,k}\log \left( 1+x_{i,j,k} \right) \right.}} \nonumber \\
&\left. \!-\!\left[ \varpi _p\lambda _p y_{i,j,k}x_{i,j,k}+\left( 1\!-\! \varpi _p \right) \Phi_{i,j,k} \right] \right\}
\end{align}
\begin{numcases}{{\rm{s.t.}}}	
t_{i,j,k}^{\mathrm{vfc}} + t_{i,j,k}^{\mathrm{A2G}} + t_{i,j,k}^{\mathrm{G2A}} \le T_{j,k}^{\max },  \label{eq:cons1} \hfill \\
{E_j}[{t_k}] - E_{i,j,k}^{\mathrm{fly}} - E_{i,j,k}^{\mathrm{A2G}} \ge  E_j^{\min }, \label{eq:cons2} \hfill \\
0 \le {y_{i,j,k}} \le y_{j,k}^{\max }. \label{eq:cons3} \hfill
\end{numcases}
{Constraint} (\ref{eq:cons1}) indicates that the total time delay of each task $\Gamma_{j,k}$ should be less than or equal to the task TTL $T_{j,k}^{\max }$. {Constraint} (\ref{eq:cons2}) implies that the remaining battery energy of UAV $j$ after task execution should not less than its minimum energy reserve $E_j^{\min }$. {Constraint} (\ref{eq:cons3}) means that the payment should be constrained by the lower bound $0$ and the upper bound $y_{j,k}^{\max }$.

\underline{Payoff function of ground vehicle.}
In VFC-based task offloading, the payoff function of each ground vehicle is associated with the payment provided by the UAV and its cost of sharing computing resources to the UAV. As such, the payoff function of ground vehicle $i$, $\forall i\in \mathcal{I}_j$ is the difference between the payment and the cost in sharing computing resource to UAV $j$, i.e.,
\begin{align}\label{eq:payoffvehicle}
{\pi _i}({x_{i,j,k}},y_{i,j,k}) &= \beta _{i,j,k}\left[\lambda _p y_{i,j,k}x_{i,j,k} - \Psi_i\left({x_{i,j,k}} \right)\right],
\end{align}
where $\Psi_i\left({x_{i,j,k}} \right)$ is the cost function, which contains two parts: the cost for sharing ${x_{i,j,k}}$ amount of computation resources and the energy consumption in data computing and transmission. According to the quadratic cost model \cite{Wu2011a}, the cost of resources sharing can be formulated as a quadratic function of the contributed resources.
As such, $\Psi_i\left({x_{i,j,k}} \right)$ can be defined as
\begin{align}\label{eq:vehiclecost}
\Psi_i\left({x_{i,j,k}} \right) = \lambda_c {\psi_i}\left({x_{i,j,k}}\right)^2 \!+\! {\lambda _e}\left (E_{i,j,k}^{\mathrm{vfc}} + P_i^{\mathrm{TX}}\, t_{i,j,k}^{\mathrm{G2A}} \right).
\end{align}
The first term in formula (\ref{eq:vehiclecost}) represents the quadratic cost in sharing ${x_{i,j,k}}$ amount of computation resources, and the second term indicates the total energy consumption in computing (i.e., $E_{i,j,k}^{\mathrm{vfc}}$) and data transmission (i.e., $P_i^{\mathrm{TX}} t_{i,j,k}^{\mathrm{G2A}}$). Here, $\lambda_c $ and ${\lambda _e}$ are positive adjustment parameters. ${\psi_i}$ is the unit cost of vehicle $i$'s computation resource.

The target of each vehicle $i\in \mathcal{I}_j$ is to maximize its payoff function by deciding the optimal AoCR strategy ${{\bf{x}}_{j}}$, whereby its optimization problem $P2$ is formulated as:
\begin{align}\label{eq:payoffvehicle2}
\mathop {\max }\limits_{x_{i,j,k}}\,&{\pi _i}({x_{i,j,k}},y_{i,j,k}) =\beta _{i,j,k}\left\{ \lambda _p y_{i,j,k}x_{i,j,k} \right. \nonumber \\
&\left. - \left[\lambda_c {\psi_i}\left({x_{i,j,k}}\right)^2 \!+\! {\lambda _e}\left (E_{i,j,k}^{\mathrm{vfc}} + P_i^{\mathrm{TX}}\; t_{i,j,k}^{\mathrm{G2A}} \right)\right] \right\}
\end{align}\vspace{-6.5mm}
\begin{align}\label{eq:constr4}
{\rm{s.t.}}~0 \le {x_{i,j,k}} \le x_{i}^{\max }.
\end{align}
{Constraint} (\ref{eq:constr4}) means that the AoCR of vehicle $i$ in executing task $\Gamma_{j,k}$ should be constrained by the lower bound $0$ and the upper bound $x_{i}^{\max }$.

\vspace{-1mm}
\subsection{Static Stackelberg Game Analysis}\label{subsec:staticgame}
During VFC-based offloading process, both UAVs and ground vehicles are rational and selfish and they are aimed to maximize their own profits \cite{ToN8884665}. Specifically, the UAV intends to enjoy high amount of computing resource with low payments, while each ground vehicle hopes the payment can be as high as possible. Based on the existing work \cite{7275166}, the competitive interactions between a UAV and ground vehicles can be formulated as a one-leader and multiple-followers Stackelberg game, i.e.,
\begin{align}\label{eq:game}
\mathbb{G} = \left\{ {\left( {j,1,2,\cdots,I_j} \right);\left( \mathbf{x}_j, \mathbf{y}_j \right);\left( {{\pi_j},{\pi_{1 \le i \le  {{I}_j} }}} \right)} \right\}.
\end{align}

In the two-stage game $\mathbb{G}$, the UAV $j$ first determines its optimal payment strategy $\mathbf{y}_j^*$ in stage I to optimize its payoff, and then each vehicle $i\in \mathcal{I}_j$ selects its optimal AoCR strategy ${x}_{i,j,k}^*$ in stage II to maximize its benefit. Here, $\mathbf{y}_j^* = \left( {y}_{i,j,k}^*\right)_{1\leq k \leq {K_j}}$ is a solution of problem $P1$, and $\mathbf{x}_j^* = \left( {x}_{i,j,k}^*\right)_{1\leq i \leq {I_j}}$ is a solution of problem $P2$. The solution of game $\mathbb{G}$ is the Stackelberg equilibrium (SE) defined in {the following} Definition~\ref{definition2}, which means neither the UAV nor the ground vehicles can improve their payoffs by deviating it.

\begin{definition}\label{definition2}
The point $\left(\mathbf{x}_j^*, \mathbf{y}_j^*\right)$ is a SE of the proposed game $\mathbb{G}$, if the following conditions hold:
\begin{align}
&~~~~\ \pi _j\left(\mathbf{x}_j^*, \mathbf{y}_j^*\right) \ge \pi _j\left(\mathbf{x}_j^*, \mathbf{y}_j\right),\label{eq:SE1}\\
&{\pi _i}({x_{i,j,k}^*},y_{i,j,k}^*) \ge {\pi _i}({x_{i,j,k}},y_{i,j,k}^*), \forall i \in {{\cal I}_j}.\label{eq:SE2}
\end{align}
\end{definition}

In the static Stackelberg game with one interaction, the parameters of the game are public known by all players.
To find the SE of the static Stackelberg game, the backward induction method is employed.
Specifically, we first investigate the decision process of every follower (i.e., vehicle) in stage II to attain its optimal AoCR strategy. Then, we analyze the optimal payment strategy of the leader (i.e., UAV) in stage I.
Every ground vehicle $i\in \mathcal{I}_j$ determines its optimal AoCR strategy ${x_{i,j,k}^*}$ to maximize its payoff ${\pi _i}({x_{i,j,k}},y_{i,j,k})$ according to the following theorem.

\emph{\textbf{ Theorem }4: }
The optimal AoCR strategies of vehicle $i$ in performing task $k$ of UAV $j$ is
\begin{align}\label{eq:AoCR}
x_{i,j,k}^{*}=\left\{ \begin{array}{ll}
	x_{i}^{\max},&\frac{2\lambda _c\psi _ix_{i}^{\max}}{\lambda _p}\le y_{i,j,k}\le y_{j,k}^{\max};\\ [0.15cm]
	\frac{\lambda _py_{i,j,k}}{2\lambda _c\psi _i},&0<y_{i,j,k}< \frac{2\lambda _c\psi _ix_{i}^{\max}}{\lambda _p};\\[0.12cm]
	0,&y_{i,j,k}=0.\\
\end{array} \right.
\end{align}

\begin{IEEEproof}
Refer to Appendix~\ref{Appendix A}
\end{IEEEproof}

Given the optimal AoCR strategy of vehicles in Eq. (\ref{eq:AoCR}), UAV $j$ determines its optimal payment strategy $\mathbf{y}_j^*$ for tasks to maximize its payoff $\pi _j\left( \mathbf{x}_j^*,\mathbf{y}_j \right)$ based on the following theorem.

\emph{\textbf{ Theorem }5: }
The optimal payment strategy of UAV $j$ on task $k$ for vehicle $i$ is
\begin{align}\label{eq:payment}
y_{i,j,k}^{*}=\left\{ \begin{array}{ll}
	\frac{2\lambda _c\psi _ix_{i}^{\max}}{\lambda _p},&\Theta _{i,j,k}\ge 0;\\[0.1cm]
	\frac{\sqrt{\Omega_{i,j,k}}-\varpi _p\lambda _c\psi _i}{\varpi _p\lambda _p},&\Theta _{i,j,k}<0.\\
\end{array} \right.
\end{align}
where
\begin{align}
\Theta _{i,j,k}&=\rho _j\alpha _{j,k}-4\varpi _p\lambda _c\psi _ix_{i}^{\max}\left( 1+x_{i}^{\max} \right),  \\
\Omega_{i,j,k} &=\varpi _p^2\lambda _c^2\psi _i^2+\varpi _p\lambda _c\psi _i\rho _j\alpha _{j,k}.
\end{align}

\begin{IEEEproof}
Refer to Appendix~\ref{Appendix B}
\end{IEEEproof}

Based on above analysis, the SE of the Stackelberg game $\mathbb{G}$ can be derived as:
\begin{align}\label{eq:stackSE}
&\left( x_{i,j,k}^{*},y_{i,j,k}^{*} \right)  \nonumber \\
&\resizebox{0.899\hsize}{!}{$= \left\{ \begin{array}{l}
	\left( x_{i}^{\max},\frac{2\lambda _c\psi _ix_{i}^{\max}}{\lambda _p} \right),~~~~~~~~~~~~~~~~~~~~~ \Theta _{i,j,k} \ge 0;\\[0.15cm]
	\left( \frac{\sqrt{\Omega _{i,j,k}}-\varpi _p\lambda _c\psi _i}{2\lambda _c\psi _i},\frac{\sqrt{\Omega _{i,j,k}}-\varpi _p\lambda _c\psi _i}{\varpi _p\lambda _p} \right),~ \Theta _{i,j,k}<0.
\end{array} \right.$}
\end{align}

\emph{\textbf{Remark:}}
In the static Stackelberg game-based offloading process where the parameters of the game are public knowledge, both UAVs and ground vehicles can apply the above SE to decide their optimal strategies to gain maximized payoffs.

\subsection{DQN-Based Offloading for Dynamic Stackelberg Game}\label{subsec:dynamicgame}
In a realistic VFC-based offloading application, the parameters of user payoff model (e.g., satisfaction, cost, and network parameters) are usually private and the parameters of network model are time-varying \cite{ToN9003500}. As a consequence, these parameters cannot be readily available for all participants. Both UAVs and ground vehicles can conduct multiple interactions and employ reinforcement learning technologies to find the optimal payment and AoCR strategies via trials, respectively, without fully knowing the accurate parameters of the network model and payoff model.
The repeated sequential interactions between a UAV and ground vehicles can be formulated as a dynamic Stackelberg game.

\underline{DQN-based payment strategy of UAV.}
A high payment of UAV for can decrease its immediate payoff, but it stimulates more vehicles' participation and their higher AoCR contribution in the future. Therefore, the current payment strategy of the UAV influences the long-term benefits.
The payment decision of UAV $j$ in the dynamic game can be formulated as a finite Markov decision process (MDP) \cite{9035635}, and reinforcement learning algorithms can be exploited to achieve the optimal payment strategy.
In particular, UAV $j$ observes the previous AoCR sequences of corresponding ground vehicles and formulates the current system state. The state of UAV $j$ at $n$-th time slot {(or interaction)} is $\mathbf{s}_j^{(n)} = \{s_{i,j,k}^{(n)}\}_{k=1}^{K_j}$, and we have $\mathbf{s}_j^{(n)} = \mathbf{x}_j^{(n-1)}$. The action of UAV $j$ at time slot $n$ is $\mathbf{y}_j^{(n)}$. For simplicity, the feasible payments of UAV $j$ are uniformly discretized into $W$ levels, i.e., $y_{i,j,k}^{(n)} \in \mathcal{W}= \{\frac{w}{W-1}\cdot y_{j,k}^{\max}\}_{0 \le w \le W-1}$.
Let $Q(\mathbf{s}_j^{(n)},\mathbf{y}_j^{(n)} )$ denote the Q-function of UAV $j$ with state-action pair $(\mathbf{s}_j^{(n)},\mathbf{y}_j^{(n)})$, which means the expected long-term discounted reward.
Note the size of state space of UAV $j$, i.e., $(I_j)^{V}$ \cite{8006228}, increases with both the number of involved vehicles (i.e., $I_j$).
Due to the presence of large state space, traditional Q-learning approaches may suffer the curse of dimensionality and cause slow learning speed and long convergence time.

In the DQN-based pricing process, to address the curse of dimensionality and accelerate the convergence rate of Q-learning, the convolutional neural network (CNN) is exploited for efficient state space compression and Q-value estimation.
Specifically, the output of the CNN is the Q-value for each payment level, i.e.,
\begin{align}\label{eq:qfunction1}\resizebox{0.894\hsize}{!}{$
Q\left( \mathbf{s}_{j}^{(n)},\mathbf{y}_{j}^{(n)} \right) \!=\! \mathbb{E}_{\mathbf{s}_j^{(n+1)}}\left[ \pi _j\left( \mathbf{x}_{j}^{(n)},\mathbf{y}_{j}^{(n)} \right)\!+\!\varsigma _j\underset{\mathbf{y}_j'}{\max}\,Q\left( \mathbf{s}_j^{(n+1)},\mathbf{y}_j' \right) \right]\!,\! $}
\end{align}
where $\varsigma_j \in [0,1]$ is the discount factor implying the myopic view of UAV $j$ about the future return. $\mathbf{s}_j^{(n+1)}$ is the new state transited from $\mathbf{s}_j^{(n)}$ with action $\mathbf{y}_j^{(n)}$. The payoff $\pi _j( \mathbf{x}_j^{(n)},\mathbf{y}_j^{(n)} )$ represents the immediate reward of UAV $j$, which is rewritten as $\pi _j^{(n)}$.
In DQN, the value of Q-function is estimated by a nonlinear neural network function approximator realized by CNN, which includes $2$ convolutional (Conv) layers and $2$ fully connected (FC) layers. The first Conv layer involves $20$ filters and each of them has size $3 \times 3$ and stride $1$. The second Conv layer contains $40$ filters and each of them has size $2 \times 2$ and stride $1$. The widely used rectified linear unit (ReLU) \cite{Deepmind} is adopted as the activation function in both Conv layers. The first FC layer uses $180$ ReLUs, while the second FC layer has $W$ ReLUs for each participating vehicle.
The architecture parameters in CNN are summarized in Table~\ref{table2}.

\begin{table}[!t]\vspace{-0.28cm}
\begin{center}
\caption{Architecture Parameters in the CNN of UAV and ground vehicle}\label{table2}\vspace{-0.5cm}
\resizebox{\linewidth}{!}{
\begin{tabular}{|c|c|c|c|c|c|c|}
    \hline
\textbf{Layer} &\textbf{Input} &\textbf{Filter Size} &\textbf{Stride} &\textbf{\# Filters} &\textbf{Activation} &\textbf{Output}\\ \hline 
    \textbf{Conv 1} &$6 \times 6$ &$3 \times 3$ &1 &20 &ReLU &$4 \times 4 \times 20$ \\ \hline
    \textbf{Conv 2} &$4 \times 4 \times 20$ &$2 \times 2$ &1 &40 &ReLU &$3 \times 3 \times 40$ \\ \hline
    \textbf{FC 1} &$360$ &$/$ &$/$ &180 &ReLU &$180$ \\ \hline
    \textbf{\tabincell{c}{FC 2 \\of UAV}} &$180$ &$/$ &$/$ &$W$ &ReLU &$W$ \\ \hline
    \textbf{\tabincell{c}{FC 2 \\of Vehicle}} &$180$ &$/$ &$/$ &$V$ &ReLU &$V$ \\ \hline
    \end{tabular} }
\end{center}\vspace{-0.5cm}
\end{table}

To efficiently learn from past experiences and smooth {the} learning, the state sequence ${\bm{\gamma}}_j^{(n)}$ is constructed for UAV $j$ in DQN, which consists of the current state and prior $A_1$ states, i.e.,
\begin{align}\label{eq:stateseq}
{\bm{\gamma}}_j^{(n)} = \left\{\mathbf{s}_{j}^{(n-A_1)},\mathbf{s}_{j}^{(n-A_1+1)}, \cdots, \mathbf{s}_{j}^{(n)} \right\}.
\end{align}
The state sequence ${\bm{\gamma}}^{(n)}$ is then reshaped into a $6\times6$ matrix and input to the CNN. We rewrite $\pi _j( \mathbf{x}_{j}^{(n)}, \mathbf{y}_{j}^{(n)} )$ as $\pi _j^{(n)}$ for simplicity. The interaction experience that UAV $j$ learned at time slot $n$ is denoted as
\begin{align}\label{eq:experience}
{\bm{\phi}}_j^{(n)} = \left\{{\bm{\gamma}}_j^{(n)},\mathbf{y}_{j}^{(n)}, \pi _j^{(n)},{\bm{\gamma}}_j^{(n+1)}  \right\},
\end{align}
and is stored into a replay memory, denoted as $\Xi_j = \left\{{\bm{\phi}}_j^{(n-d+1)},\cdots,{\bm{\phi}}_j^{(n)} \right\}$. Here, only {the} latest $d$ related experiences are stored to save memory space and restrain UAV $j$ from focusing on the immediate interaction experience. Let ${\bm{\theta}}^{(n)}$ denote the filter weights of UAV $j$ in the CNN. Based on the experience replay method, ${\bm{\theta}}^{(n)}$ is updated for $D_1$ times at every time slot by randomly selecting an experience from $\Xi_j$ to minimize the mean-squared error of the target optimal Q-function. According to \cite{Deepmind}, the loss function can be defined as:
\begin{align}\label{eq:loss}\resizebox{1.03\hsize}{!}{$
\mathscr{L}( {\bm{\theta}}^{(n)} ) \!=\!\mathbb{E}_{{\bm{\phi}}_{j}^{(n)}}\left[ \left( \pi _j^{(n)} \!+\!\varsigma _j\underset{\mathbf{y}_j'}{\max}\,Q( {\bm{\gamma}} _{j}^{( n+1)},\mathbf{y}_j';{\bm{\theta}} ^{( n-1 )} )  \!-\!Q( \gamma _{j}^{(n)},\mathbf{y}_{j}^{(n)};{\bm{\theta}}^{(n)} ) \right) ^2 \right]\!.\!$}
\end{align}
The stochastic gradient descent (SGD) algorithm is adopted to update ${\bm{\theta}}^{(n)}$ for alleviated computational cost in learning via mini-batch updates.
Based on the CNN model and current system state $\mathbf{s}_j^{(n)}$, UAV $j$ applies the $\epsilon$-greedy policy to choose its action $\mathbf{y}_j^{(n)}$ for a better tradeoff between exploration and exploitation. More specifically, the greedy payment vector that maximizes its Q function is chosen with a high probability $\epsilon_j$, and other actions are randomly selected with a very small chance $1-\epsilon_j$, i.e.,
\begin{align}\label{eq:epsilon1}
\Pr\left[\mathbf{y}_j^{(n)} \!=\! \mathbf{y}_j^*\right] \!=\! \left\{ \begin{array}{ll}
\epsilon_j, ~~~~\;\mathbf{y}_j^* \!=\! \arg {\max \limits_{\mathbf{y}_j'}}\,Q( \mathbf{s}_j^{(n)}, {\mathbf{y}_j'});\\
{1 \!-\! \epsilon_j }, ~{\rm{otherwise}}.
\end{array} \right.
\end{align}
The detailed procedure of DQN-based payment for UAV $j$ is shown in Algorithm~\ref{Algorithm1}.

\begin{algorithm}[t!]\begin{small}
   \caption{\textbf{DQN-Based Optimal Payment Strategy}}\label{Algorithm1}
    \begin{algorithmic}[1]
        \STATE \textbf{Initialize: }${\bm{\theta}}^{(0)}$, $\varsigma_j$, $\Xi_j =\emptyset$, $\mathbf{s}_{j}^{(0)}$, $A_1$, $D_1$, $\mathcal{W}$
        \FOR {$n = 1, 2,\cdots,N $}
            \STATE Set $\mathbf{s}_j^{(n)} = \mathbf{x}_j^{(n-1)}$.
            \IF {$n \le A_1 $}
                \STATE Select $\mathbf{y}_j^{(n)} \in \mathcal{W}^{I_j}$ at random.
            \ELSE
                \STATE Build ${\bm{\gamma}}_j^{(n)} = \{\mathbf{s}_{j}^{(n-A_1)},\mathbf{s}_{j}^{(n-A_1+1)}, \cdots, \mathbf{s}_{j}^{(n)} \}$.
                \STATE Obtain CNN output $Q( \mathbf{s}_{j}^{(n)},\mathbf{y}_{j} )$, $\forall \mathbf{y}_{j}\in \mathcal{W}^{I_j}$, as Q-values with input ${\bm{\gamma}}_j^{(n)}$ and ${\bm{\theta}}^{(n)}$.
                \STATE Select $\mathbf{y}_{j}^{(n)}$ via $\epsilon$-greedy policy in Eq. (\ref{eq:epsilon1}) and send ${y}_{i,j,k}^{(n)}$ to vehicle $i\in \mathcal{I}_j$.
            \ENDIF
            \STATE Observe and evaluate the AoCR ${\bf{x}}_{j}^{(n)}$.
            \STATE Compute payoff $\pi _j( \mathbf{x}_j^{(n)},\mathbf{y}_j^{(n)} )$ via Eq. (\ref{eq:payoffUAV2}).
            \STATE Update $\Xi_j \leftarrow \{{\bm{\gamma}}_j^{(n)},\mathbf{y}_{j}^{(n)}, \pi _j^{(n)},{\bm{\gamma}}_j^{(n+1)}  \} \cup \Xi_j$.
            \FOR {$d = 1, 2,\cdots,D_1 $}
            \STATE Select $\{{\bm{\gamma}}_j^{(d)},\mathbf{y}_{j}^{(d)}, \pi _j^{(d)},{\bm{\gamma}}_j^{(d+1)} \} \in \Xi_j$ at random.
            \STATE Update ${\bm{\theta}}^{(n)}$ via mini-batch SGD.
            \ENDFOR
        \ENDFOR
    \end{algorithmic}\end{small}\vspace{-0.18mm}
\end{algorithm}

\begin{algorithm}[t!]\begin{small}
   \caption{\textbf{DQN-Based Optimal AoCR Strategy}}\label{Algorithm2}
    \begin{algorithmic}[1]
        \STATE \textbf{Initialize: }$\tilde{{\bm{\theta}}}^{(0)}$, $\varsigma_i$, $\Xi_i =\emptyset$, ${\tilde{s}}_{i,j,k}^{(0)}$, $A_2$, $D_2$, $\mathcal{V}$
        \FOR {$n = 1, 2,\cdots,N $}
            \FOR {$i = 1, 2,\cdots,I_j $}
                \STATE Set ${\tilde{s}}_{i,j,k}^{(n)} = {y}_{i,j,k}^{(n-1)}$.
                \IF {$n \le A_1 $}
                    \STATE Select ${x}_{i,j,k}^{(n)} \in \mathcal{V}$ at random.
                \ELSE
                    \STATE Build ${\bm{\gamma}}_i^{(n)} = \{{\tilde{s}}_{i,j,k}^{(n-A_2)},{\tilde{s}}_{i,j,k}^{(n-A_2+1)}, \cdots, {\tilde{s}}_{i,j,k}^{(n)}\}$.
                    \STATE Obtain CNN output $\tilde{Q}({\tilde{s}}_{i,j,k}^{(n)},{x}_{i,j,k})$, $\forall {x}_{i,j,k}\in \mathcal{V}$, as Q-values with input ${\bm{\gamma}}_i^{(n)}$ and $\tilde{{\bm{\theta}}}^{(n)}$.
                    \STATE Select ${x}_{i,j,k}^{(n)}$ via $\epsilon$-greedy policy in Eq. (\ref{eq:epsilon2}) and send it to UAV $j$.
                \ENDIF
                \STATE Observe and evaluate the payment ${y}_{i,j,k}^{(n)}$.
                \STATE Compute payoff $\pi _i({x}_{i,j,k}^{(n)}, {y}_{i,j,k}^{(n)})$ via Eq. (\ref{eq:payoffvehicle2}).
                \STATE Update $\Xi_i \leftarrow \{{\bm{\gamma}}_i^{(n)},{x}_{i,j,k}^{(n)}, \pi _i^{(n)},{\bm{\gamma}}_i^{(n+1)} \} \cup \Xi_i$.
                \FOR {$d = 1, 2,\cdots,D_2 $}
                \STATE Select $\{{\bm{\gamma}}_i^{(d)},{x}_{i,j,k}^{(d)}, \pi _i^{(d)},{\bm{\gamma}}_i^{(d+1)}\} \in \Xi_i$ at random.
                \STATE Update $\tilde{{\bm{\theta}}}^{(n)}$ via mini-batch SGD.
                \ENDFOR
            \ENDFOR
        \ENDFOR
    \end{algorithmic}\end{small}\vspace{-0.18mm}
\end{algorithm}

\underline{DQN-based AoCR strategy of ground vehicle.}
The AoCR decision of each ground vehicle can be formulated as an MDP with finite states. As it is unrealistic for vehicles to have full knowledge of the network model and the UAV's private payoff model in time, each vehicle $i$ can utilize the DQN method to obtain its optimal AoCR strategy in the dynamic game.
In the DQN-based AoCR decision-making process, the payment of UAV $j$ at the last time slot is used by vehicle $i$ as the current state to decide the current AoCR action. The system state of vehicle $i$ at time slot $n$ is ${\tilde{s}}_{i,j,k}^{(n)} = y_{i,j,k}^{(n-1)}$, and its action at time slot $n$ is ${x}_{i,j,k}^{(n)}$. For simplicity, the feasible AoCRs of vehicle $i$ are discretized into $V$ levels, i.e., $x_{i,j,k}^{(n)} \in \mathcal{V}= \{\frac{v}{V-1}\cdot x_{i}^{\max}\}_{0 \le v \le V-1}$.
The Q-function of vehicle $i$ for every state-action pair $({\tilde{s}}_{i,j,k}^{(n)},{x}_{i,j,k}^{(n)})$ is denoted by $\tilde{Q}({\tilde{s}}_{i,j,k}^{(n)},{x}_{i,j,k}^{(n)})$ and is estimated by the CNN, i.e.,
\begin{align}\label{eq:qfunction2}\resizebox{0.894\hsize}{!}{$
\tilde{Q}\left({\tilde{s}}_{i,j,k}^{(n)},{x}_{i,j,k}^{(n)}\right) \!=\! \mathbb{E}_{{\tilde{s}}_{i,j,k}^{(n+1)}}\left[ \pi _i^{(n)} \!+\! \varsigma _i \underset{{x}_{i,j,k}'}{\max}\,\tilde{Q}\left( {\tilde{s}}_{i,j,k}^{(n+1)},{x}_{i,j,k}' \right) \right]\!,\! $}
\end{align}
where $\varsigma_i \in [0,1]$ is the discount factor implying the myopic view of vehicle $i$ about the future return. ${\tilde{s}}_{i,j,k}^{(n+1)}$ is the new state transited from ${\tilde{s}}_{i,j,k}^{(n)}$ with action ${x}_{i,j,k}^{(n)}$. The payoff $\pi _i({x}_{i,j,k}^{(n)},{y}_{i,j,k}^{(n)})$ means the immediate reward of vehicle $i$, and is rewritten as $\pi _i^{(n)}$ for simplicity.
As shown in Table~\ref{table2}, the parameters of CNN for each vehicle are similar to those of UAV $j$ except that this CNN has $V$ outputs.

Each vehicle $i$ builds its state sequence ${\bm{\gamma}}_i^{(n)}$ in DQN and reshapes it into a $6\times6$ matrix as the input of the CNN. The state sequence includes the recent $A_2+1$ states, i.e.,
\begin{align}\label{eq:stateseq2}
{\bm{\gamma}}_i^{(n)} = \left\{{\tilde{s}}_{i,j,k}^{(n-A_2)},{\tilde{s}}_{i,j,k}^{(n-A_2+1)}, \cdots, {\tilde{s}}_{i,j,k}^{(n)} \right\}.
\end{align}
Then, each vehicle $i$ stores its previous interaction experiences into a replay memory $\Xi_i = \left\{{\bm{\phi}}_i^{(n-d+1)},\cdots,{\bm{\phi}}_i^{(n)} \right\}$, where ${\bm{\phi}}_i^{(n)}$ is the experience that vehicle $i$ has acquired at time slot $n$, i.e.,
\begin{align}\label{eq:experience2}
{\bm{\phi}}_i^{(n)} = \left\{{\bm{\gamma}}_i^{(n)},{x}_{i,j,k}^{(n)},\, \pi _i^{(n)},{\bm{\gamma}}_i^{(n+1)}  \right\}.
\end{align}
The parameters of the Q-network of vehicle $i$, denoted by $\tilde{{\bm{\theta}}}^{(n)}$, is updated by SGD based on the loss function, which is similar to Eq. (\ref{eq:loss}).
The SGD training process is repeated $D_2$ times to update $\tilde{{\bm{\theta}}}^{(n)}$ at each time slot based on the randomly chosen experience from the replay memory, i.e., ${\bm{\phi}}_i^{(d)}\in \Xi_i$.

To avoid staying at the local optimum, the $\epsilon$-greedy policy is exploited by each vehicle $i$ to opt its action ${x}_{i,j,k}^{(n)}$, i.e.,
\begin{align}\label{eq:epsilon2}\resizebox{0.899\hsize}{!}{$
\Pr\left[{x}_{i,j,k}^{(n)} \!=\! {x}_{i,j,k}^*\right] \!=\! \left\{ \begin{array}{ll}
\epsilon_i, ~~~~{x}_{i,j,k}^* \!=\! \arg {\max \limits_{{x}_{i,j,k}'}}\,\tilde{Q}( \tilde{{s}}_{i,j,k}^{(n)}, {x}_{i,j,k}');\\
{1 \!-\! \epsilon_i }, ~{\rm{otherwise}}.
\end{array} \right.$}
\end{align}
The detailed procedure of DQN-based AoCR for vehicle $i\in \mathcal{I}_j$ is shown in Algorithm~\ref{Algorithm2}.

\emph{\textbf{Remark:}} The proposed two-tier DQN algorithm satisfies both convergency and robustness in the VFC-based offloading process.

\emph{1) Convergency}.
According to \cite{Watkins2004NQ}, it is theoretically guaranteed that Q-learning can converge to the optimal action-value function when $n \rightarrow \infty$.
Besides, based on the experimental validations in \cite{Deepmind}, the large CNN can be efficiently trained in a stable manner in DQN via experience replay and SGD. The stability of CNN training and the convergency of the two-tier DQN algorithm are demonstrated in the simulation section.

\emph{2) Robustness}.
By incorporating CNNs for state space compression and experience replay methods for mitigation of oscillations or divergence in learning, our DQN-based offloading algorithm can be effective in a large-scale network involving a large number of vehicles \cite{8006228}.
Besides, according to the experimental results in \cite{8310016}, the training data obtained from the historical experiments performed in similar scenarios can be exploited to accelerate the learning speed of Q-learning. Hence, each UAV or vehicle can pre-train its DQN model in an off-line manner {(e.g., when they are in recharging state)} by exploiting previous experiences learned from similar scenarios to speed up the convergence rate.

\section{PERFORMANCE EVALUATION}\label{sec:SIMULATION}
\subsection{Simulation Setup}

\begin{figure}[t!]\setlength{\abovecaptionskip}{-0.05cm}
\centering
\setlength{\abovecaptionskip}{-0.02cm}
  \includegraphics[width=9.4cm]{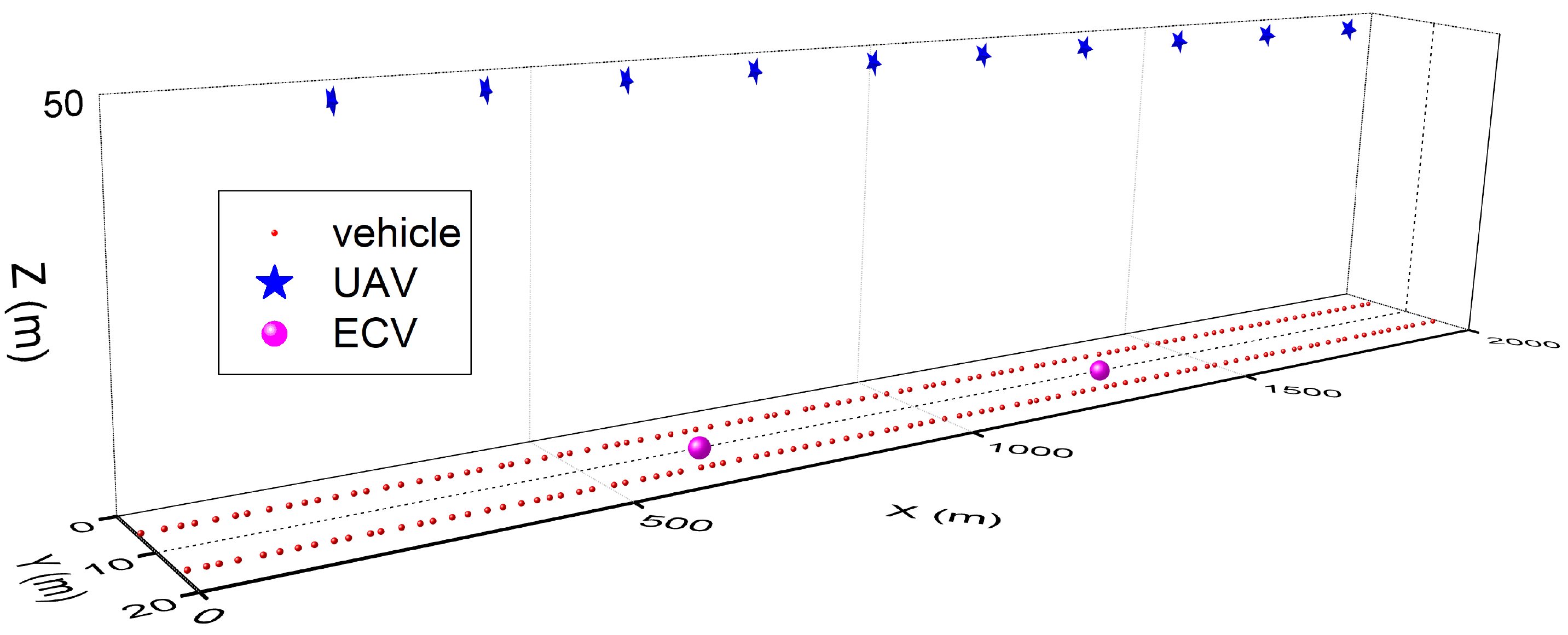}
  \caption{The simulation area.}\label{fig:simulation}\vspace{-0.2cm}
\end{figure}

As shown in Fig~\ref{fig:simulation}, we consider a simulation scenario with $10$ UAVs, $200$ ground vehicles, and $2$ ECVs in a disaster site with width of $20$m and length of $2000$m.
UAVs fly along the predefined straight-line trajectory \cite{Liu2019tran} with a fixed altitude of $50$m over the two-lane road.
UAVs are initially deployed alongside the road every $200$m, and vehicles are randomly distributed on each lane of the road with minimum safe distance of $10$m.
The minimum and maximum velocities of ground vehicles on the road are set as $24$ km/h and $72$ km/h.
The number of UAV's offloading tasks follows the uniform distribution between $10$ and $20$. The data size of each task $D_{j,k}$ is randomly in $[1, 10]$ Mbits. The required CPU cycles $\theta_{i,j,k}$ for one-bit data processing is randomly selected in $[100,200]$ CPU cycles/bit \cite{8908666}.
Each vehicle's cost parameter ${\psi_i}$ is uniformly distributed between $4$ and $16$ cents.
Our RescueChain system is established atop the Tendermint Core consensus engine \cite{githubTendermint} by utilizing the Docker container environment run on three computers with Intel Core i7-8700 CPU, 8GB RAM, and Ubuntu 16.04 OS.
Other parameters are summarized in Table \ref{table3}.

\begin{table}[!t]
    \begin{center}
        \caption{Simulation Parameters}\label{table3}\vspace{-0.2cm}
        \begin{tabular}{cc||cc}
        \hline \hline
        \textbf{Parameter} & \textbf{Value} & \textbf{Parameter} & \textbf{Value}  \\ \hline
            $v_j^{\max} $&$20$ m/s \cite{8908666} &$ P_i^{\mathrm{TX}} $&$0.1$ W \cite{8956055}  \\[1.7pt] 
            $ P_j^{\mathrm{TX}} $&$1$ W \cite{8956055} &$ P_j^{\mathrm{RX}} $&$0.1$ W \cite{8956055}  \\ 
            $ {\varphi _0} $&$-50$ dBm \cite{8468999} &$ \sigma_0^2 $&$-100$ dBm \cite{8956055} \\[3pt] 
            $ B_{i}^{\mathrm{UL}} $&$10$ MHz \cite{8956055} &$ B_{j}^{\mathrm{DL}} $&$0.5$ MHz \cite{8956055} \\[2.5pt] 
            $ R_j^{\mathrm{A2A}}$&$400$ m &$ R_j^{\mathrm{A2G}}$&$200$ m  \\[-0.2pt]
            $ \mu $&$2$ \cite{9035635} &$ \kappa_i $&$10^{-28}$ \cite{8908666} \\
            $ \lambda_1,\lambda_2 $&$0.0037,5.0206$\cite{7932157TVT} &$C_j  $&$500$ kJ \cite{7932157TVT} \\ 
            $ {\rho _j}, a_j $&$162,2\,\mathrm{m/s^2}$ &$ {\varpi_p} $&$0.5$  \\ 
            $ {\lambda_p},\lambda_c $&$8,0.05$ &${\alpha _{j,k}}  $&$[0.1,0.9]$  \\[1pt] 
            $ y_{j,k}^{\max }$&$11$ cents &$ x_{i}^{\max }$&$6$ GHz  \\ 
            $ \varsigma_j, \varsigma_i $&$0.8$ &$ \epsilon_j,\epsilon_i $&$0.92,0.95$  \\
            $ A_1, A_2 $&$11$ &$D_1, D_2 $&$4$  \\ 
            $ \left|\Xi_j \right|, \left|\Xi_i \right| $&$1000$ &$W, V $&$22,12$  \\ 
            $ N $&$9000$ &$ \theta_{j,k}$&$[0.3,0.7]$  \\ 
            $ Z, \Psi $&$10,7$ &$\Im_u^{\mathrm{ini}},\Delta_{vol} $&$3$  \\ 
            $\Delta_{cp}, \Delta_{cv} $&$5,3$ &$\Delta_{wbc}, \Delta_{nbc} $&$5,1.5$  \\ 
            $\Delta_{sbc},\Delta_{sbv} $&$4,2$ &$ \Delta_{rep},\Delta_{acc} $&$1.5,2.5$  \\
            \hline \hline
        \end{tabular}
    \end{center}\vspace{-0.47cm}
\end{table}

\begin{figure*}[!t]\setlength{\abovecaptionskip}{-0.05cm}\vspace{-0.3cm}
\begin{minipage}[t]{0.315\textwidth}
\centering
    \includegraphics[height=4.0cm,width=\textwidth]{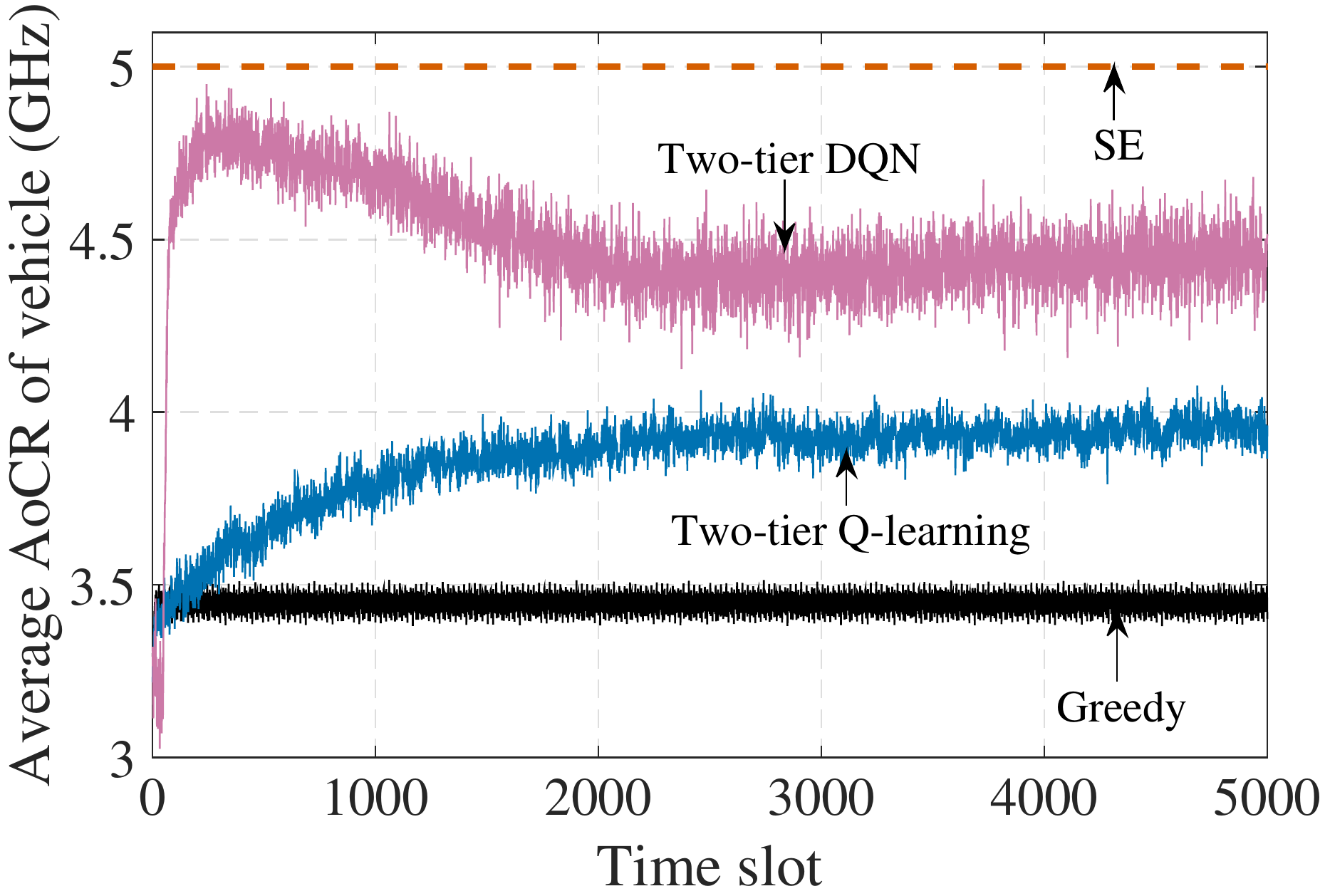}
    \caption{Evolution of average AoCR of vehicles, compared with two existing schemes.}\label{fig:simu1}
\end{minipage}~~~~
\begin{minipage}[t]{0.315\textwidth}
\centering
    \includegraphics[height=4.0cm,width=\textwidth]{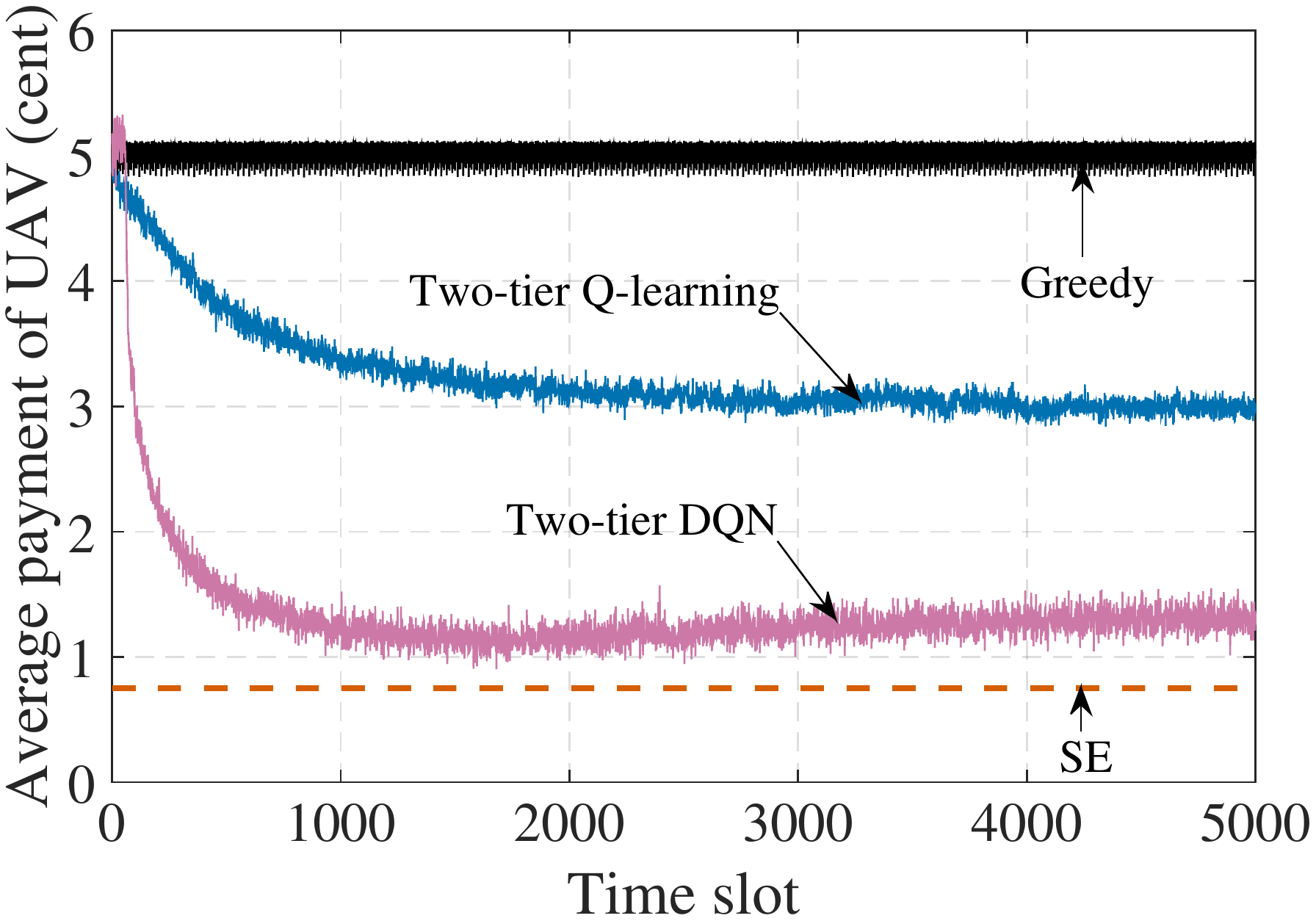}
    \caption{Evolution of average payment of UAVs, compared with two existing schemes.}\label{fig:simu2}
\end{minipage}~~~~
\begin{minipage}[t]{0.315\textwidth}
\centering
    \includegraphics[height=4.0cm,width=\textwidth]{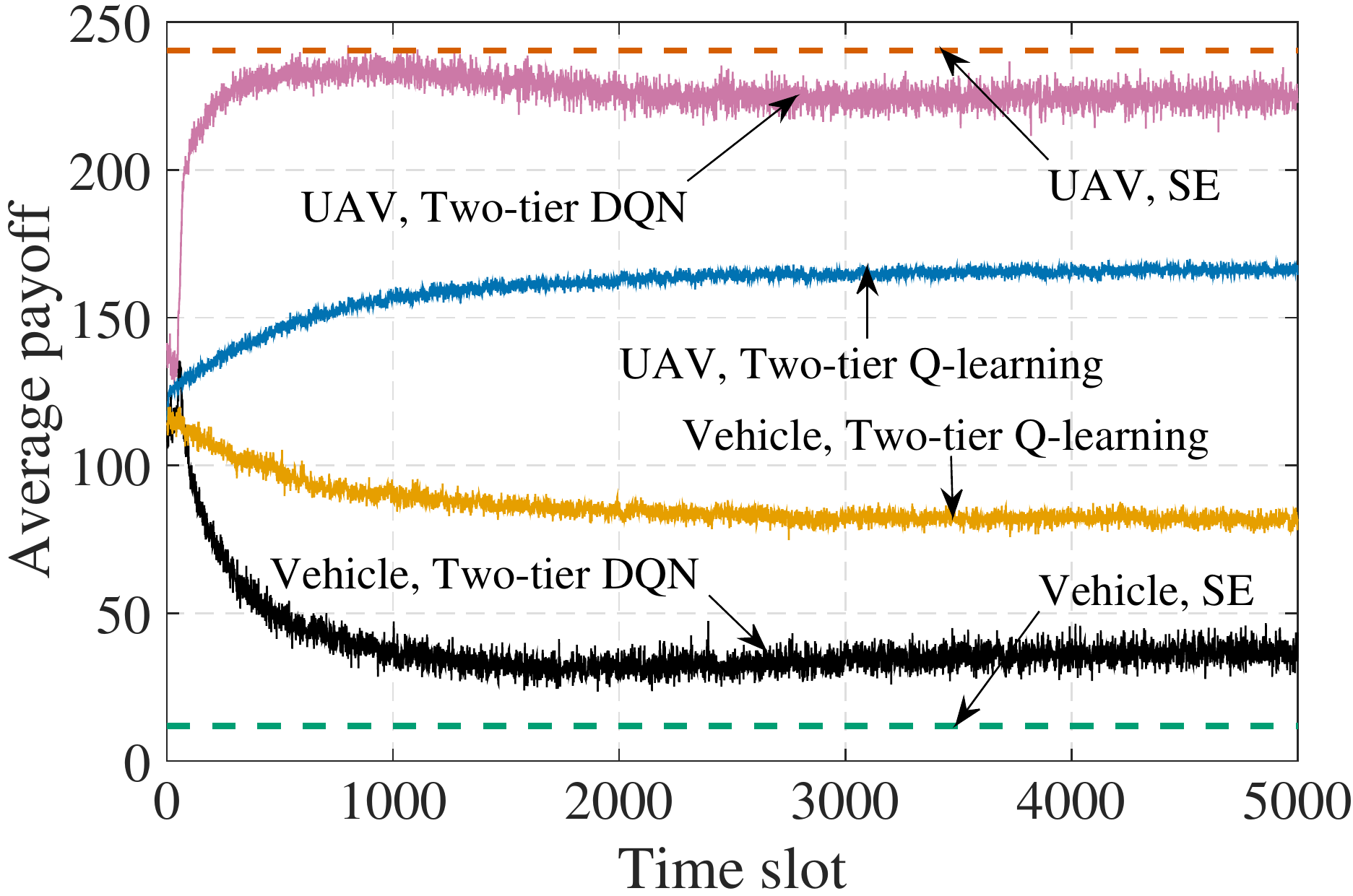}
    \caption{Evolution of average payoff of UAVs and vehicles, compared with two-tier Q-learning scheme.}\label{fig:simu3}
\end{minipage}
\end{figure*}

\begin{figure*}[!tbp]\setlength{\abovecaptionskip}{-0.05cm}\vspace{-0.2cm}
\begin{minipage}[t]{0.313\textwidth}
\centering
    \includegraphics[height=4.07cm,width=0.96\textwidth]{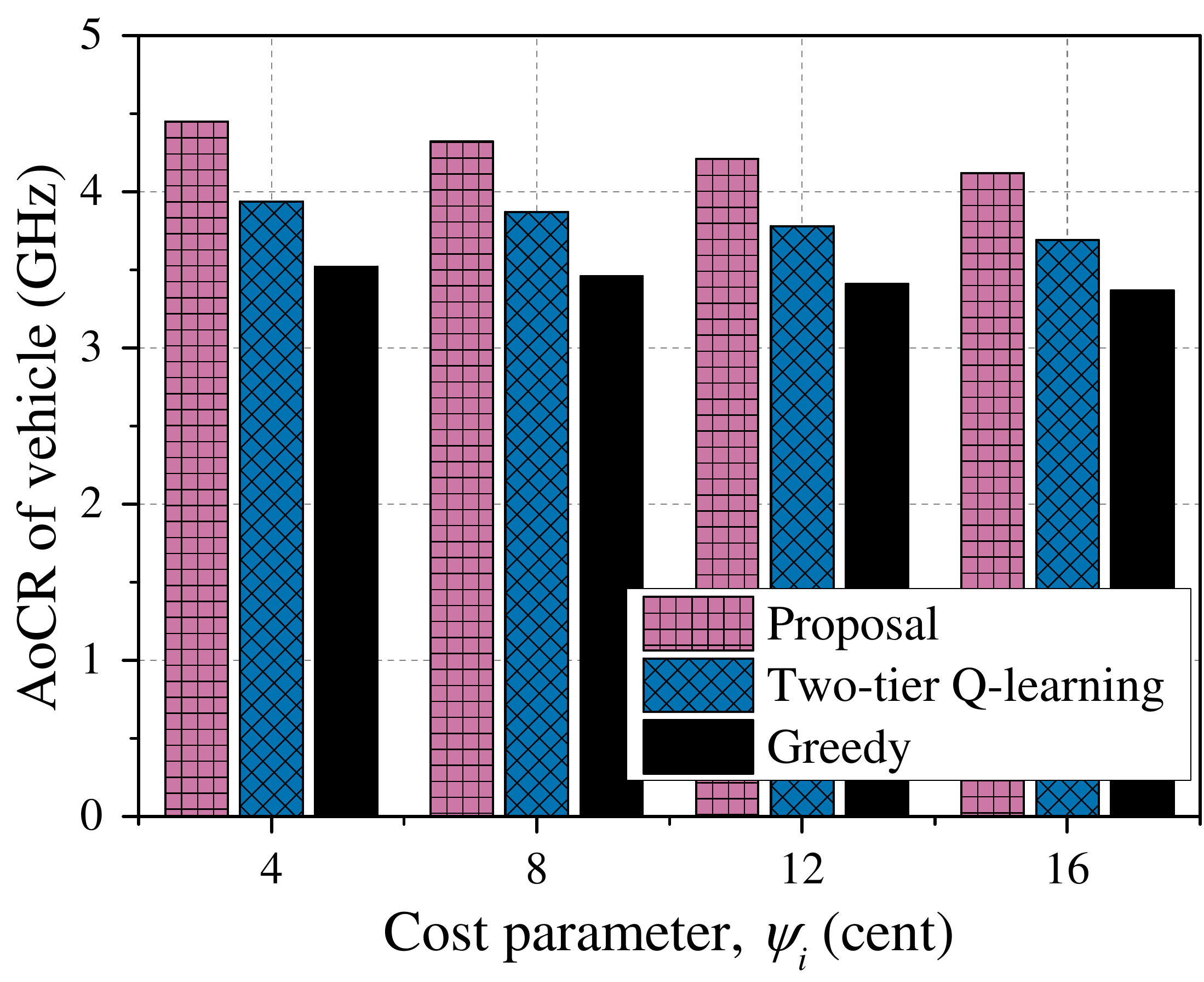}
    \caption{Optimal AoCR of vehicle vs. cost parameter $\psi_i$, compared with two existing schemes.}\label{fig:simu4}
\end{minipage}~~~~
\begin{minipage}[t]{0.313\textwidth}
\centering
    \includegraphics[height=4.07cm,width=0.96\textwidth]{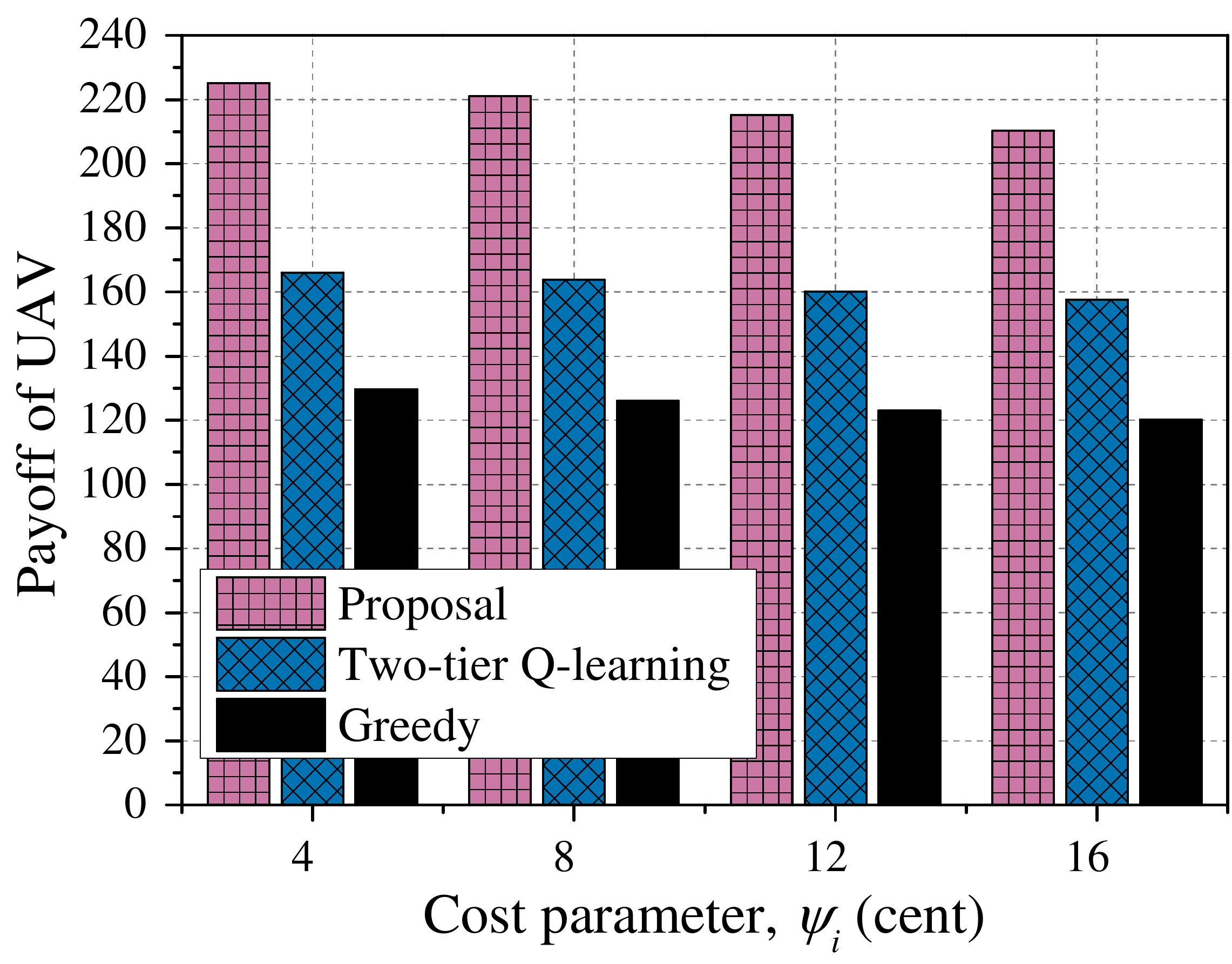}
    \caption{Payoff of UAV vs. cost parameter $\psi_i$, compared with two existing schemes.}\label{fig:simu5}
\end{minipage}~~~~~~
\begin{minipage}[t]{0.315\textwidth}
\centering
    \includegraphics[height=4.07cm,width=\textwidth]{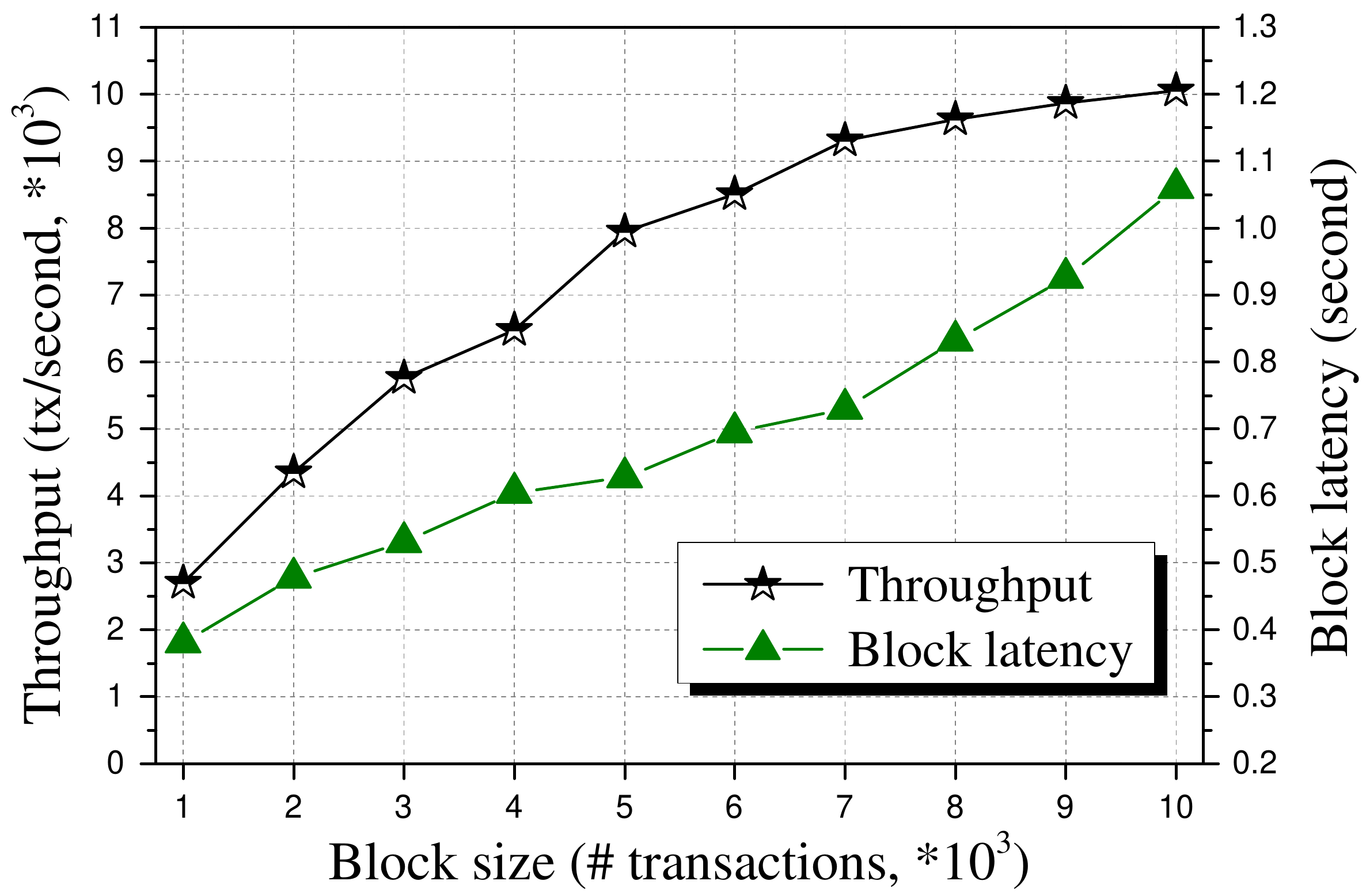}
    \caption{Throughput and block latency vs. block size in RescueChain.}\label{fig:simu6}
\end{minipage}\vspace{-0.3cm}
\end{figure*}

In the simulation, we consider two kinds of \emph{insider attacks} launched by malicious insider UAVs: (i) \emph{spoofing attack}: they behave honestly at first to gain reputation and then begin to commit misbehaviors; and (ii) \emph{collusion attack}: they collude to vote malicious delegates and witness incorrect blocks in the consensus process.
The performance of our proposal is evaluated by comparing with the following conventional schemes:
\begin{itemize}
  \item \textit{Two-tier Q-learning scheme \cite{Wang2021INFOCOM}:} both ground vehicles and UAVs employ the Q-learning method to seek their optimal AoCR and pricing strategies in the dynamic Stackelberg game, respectively.
  \item \textit{Greedy scheme:} both ground vehicles and UAVs greedily opt their optimal AoCR and pricing strategies during repeated interactions in the dynamic Stackelberg game, respectively.
  \item \textit{Accumulative reputation-based Tendermint (ART) scheme:} the reputation of validators in Tendermint are evaluated using the accumulative behavior effects in Eq. (\ref{4-credit}) with $\eta=0$ and $\Psi=Z$, where the level-2 validators and the time decay effect are not taken into account.
  \item \textit{Naive Tendermint scheme \cite{Buchman2016Tendermint}:} the naive Tendermint protocol is operated by all full nodes to achieve consensus without considering the reputation assessment of full nodes.
\end{itemize}

\vspace{-0.1cm}
\subsection{Simulation Results}

\begin{figure*}[!tbp]\setlength{\abovecaptionskip}{-0.05cm}\vspace{-0.2cm}
\begin{minipage}[t]{0.31\textwidth}
\centering
    \includegraphics[height=4.07cm,width=\textwidth]{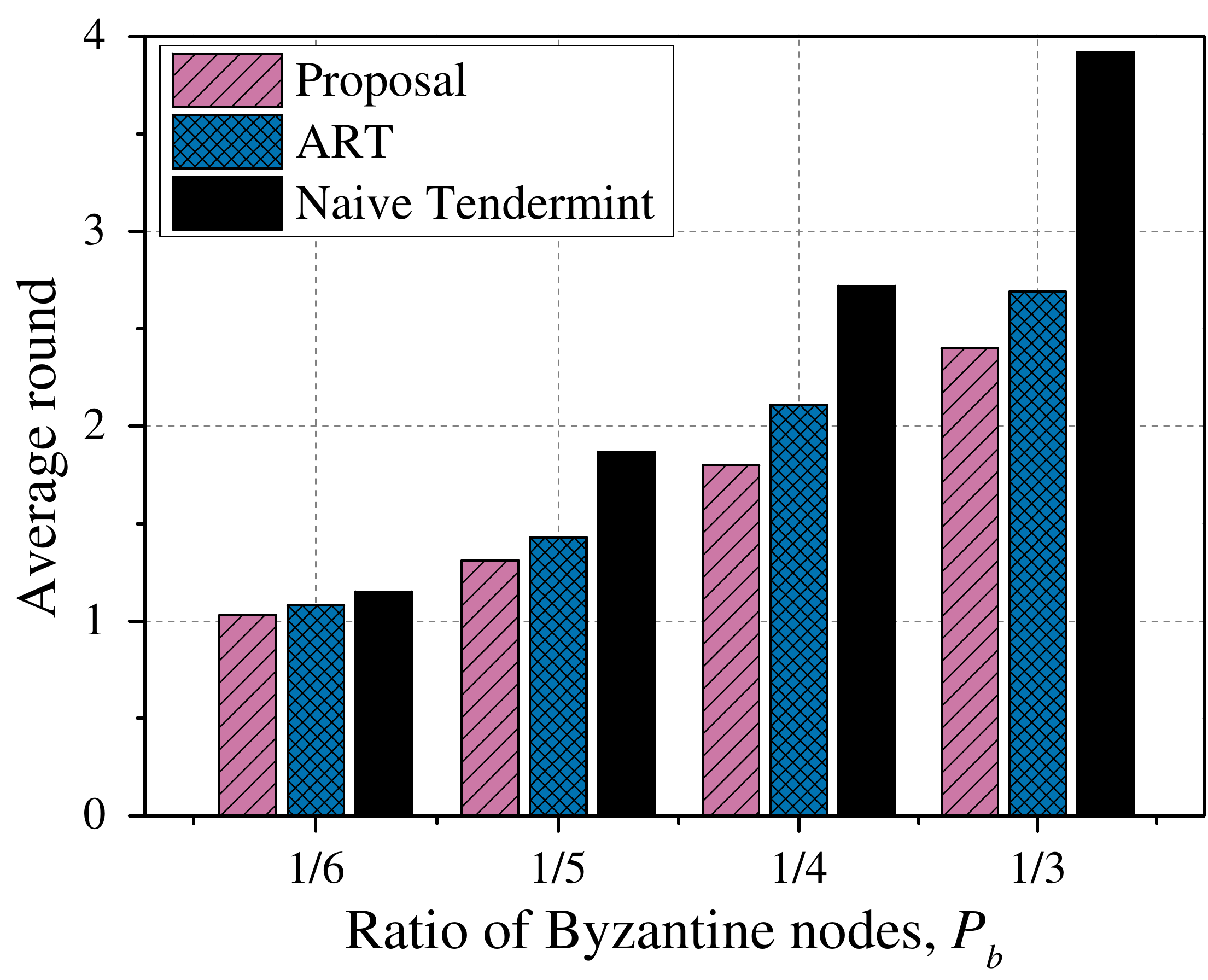}
    \caption{Average round to reach consensus vs. ratio of Byzantine nodes $P_b$, compared with two existing schemes.}\label{fig:simu7}
\end{minipage}~~~~~
\begin{minipage}[t]{0.31\textwidth}
\centering
    \includegraphics[height=4.07cm,width=\textwidth]{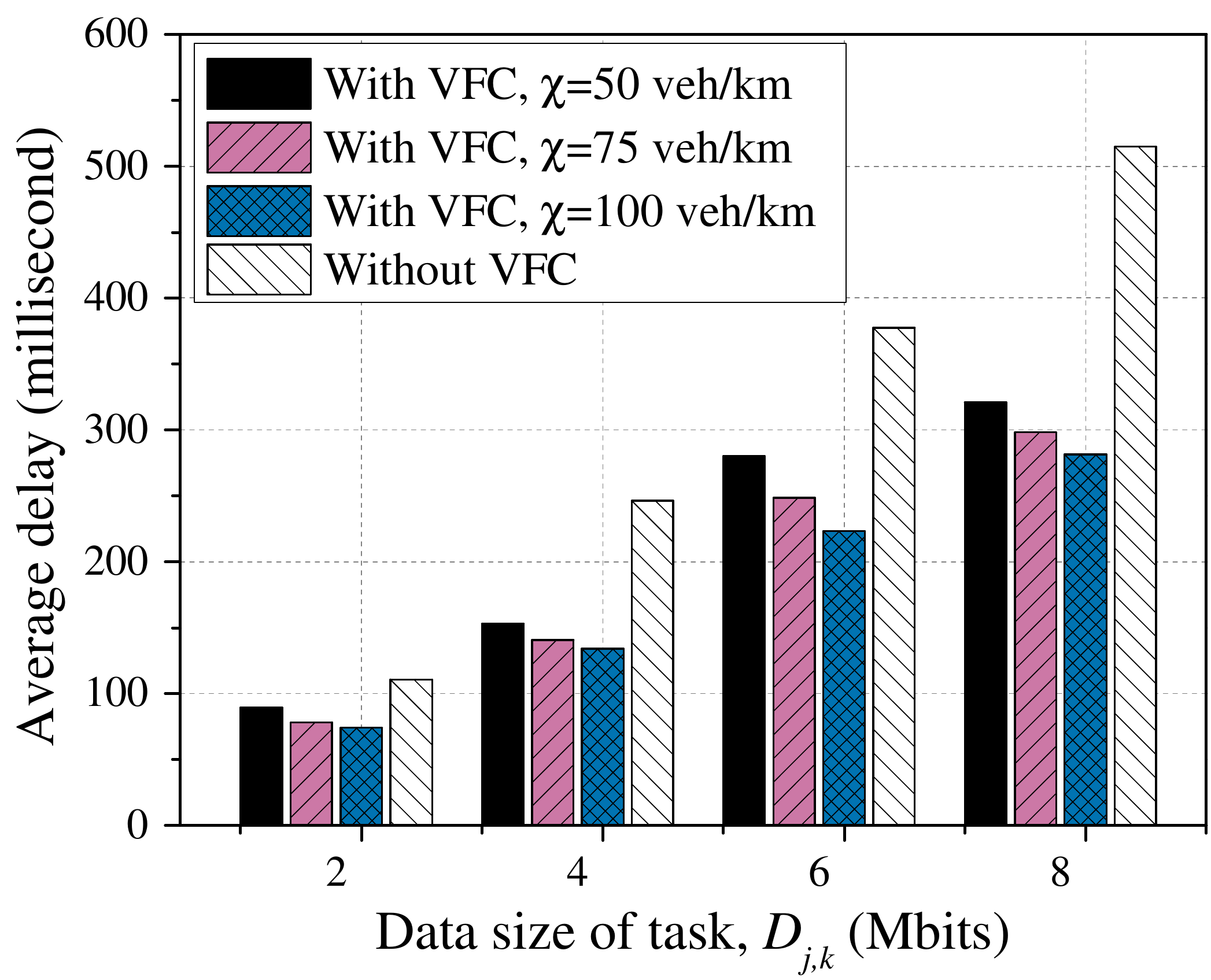}
    \caption{Average delay in offloading vs. data size ${D_{j,k}}$ of task with different vehicle density $\chi$, compared with the scheme without VFC.}\label{fig:simu8}
\end{minipage}~~~~~
\begin{minipage}[t]{0.31\textwidth}
\centering
    \includegraphics[height=4.07cm,width=\textwidth]{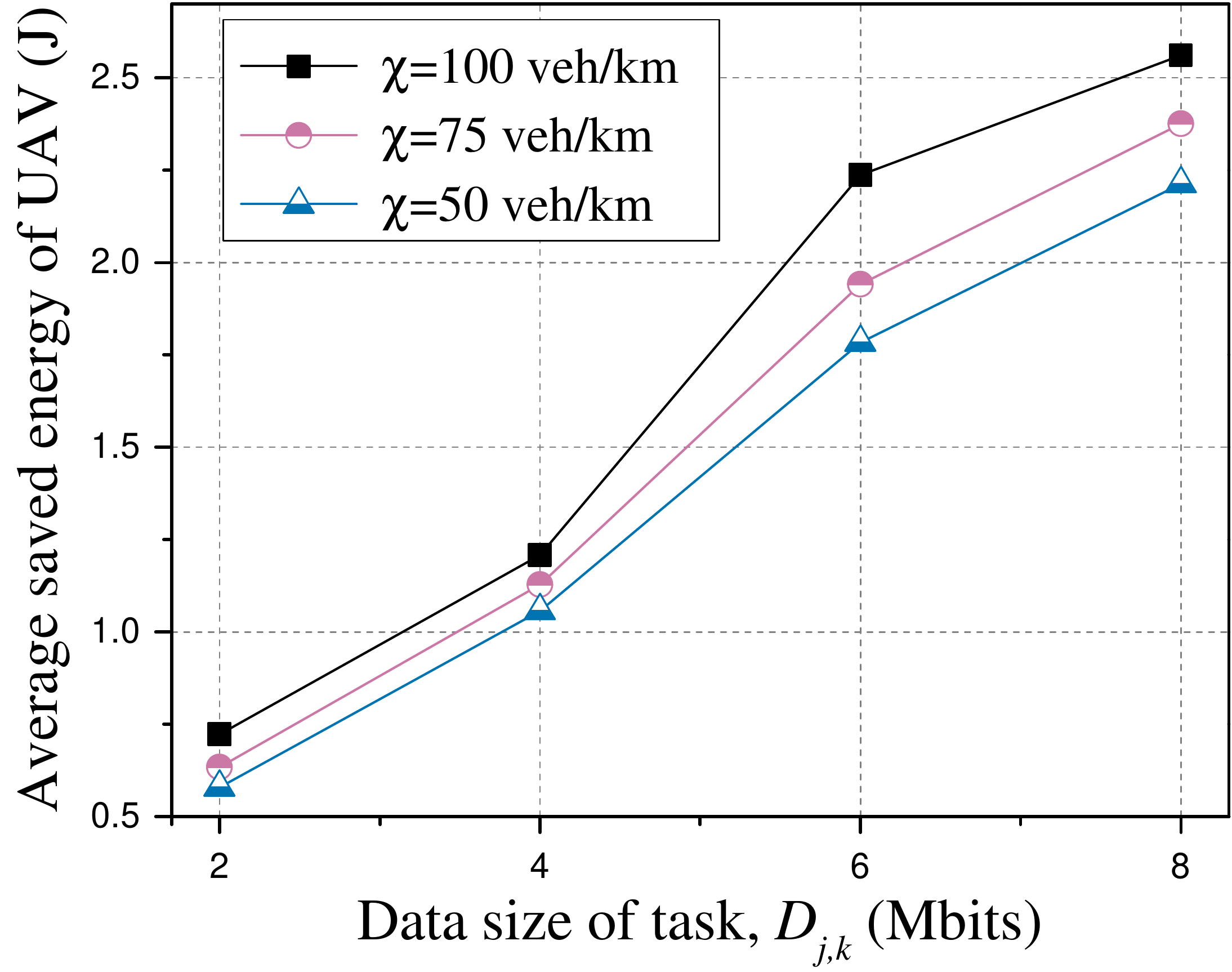}
    \caption{Average saved energy of UAV in offloading vs. data size ${D_{j,k}}$ of task with different vehicle density $\chi$.}\label{fig:simu9}
\end{minipage}\vspace{-0.3cm}
\end{figure*}

Fig.~\ref{fig:simu1} and Fig.~\ref{fig:simu2} show the evolutions of average AoCR and payment of ground vehicles and UAVs in different schemes. As seen in the two figures, the proposed two-tier DQN scheme effectively motivates higher AoCR of vehicles with lower payment and attains a smaller deviation with the SE analyzed in Eq. (\ref{eq:stackSE}), compared with the two-tier Q-learning scheme and the greedy scheme. It can be explained as follows. In the two-tier Q-learning scheme, owing to the curse of dimensionality in searching the large state space, it is difficult for the UAV to efficiently find its optimal payment policy via Q-learning to encourage vehicles' high computing resource contributions, thereby resulting in a relatively lower AoCR and higher payment. In the greedy scheme, since both UAVs and ground vehicles behave greedily during the strategy-making process, they can only attain the local optimum of the payoff, causing the low AoCR and high payment.
Besides, the AoCR in our two-tier DQN scheme in Fig.~\ref{fig:simu1} first increases then decreases, and finally reaches a stable value. The reason is that the initial high payment of UAV can motivate vehicles' high AoCR, while vehicles tend to decrease the AoCR after observing the subsequent decreasing payment.

Fig.~\ref{fig:simu3} show the evolutions of average payoffs of UAVs and vehicles in three schemes. From Fig.~\ref{fig:simu3}, the proposed two-tier DQN scheme outperforms the two-tier Q-learning scheme in attaining a smaller deviation with the SE, as our proposal can promote vehicles' higher AoCR contributions with lower payments as shown in the above two figures.
Besides, as time grows, the average payoff of UAVs increases, while that of vehicles is in a decline. It is because after seeing the initial high AoCR of vehicles, the UAV prefers decreasing its payment gradually to improve its payoff. Meanwhile, after observing UAV's prior payments, vehicles intend to increase their AoCR to seek optimized profit.

Fig.~\ref{fig:simu4} and Fig.~\ref{fig:simu5} compare the proposed scheme with other two schemes in terms of the average AoCR of vehicles and the average payoff of UAVs, where the cost parameter $c_i$ of vehicle varies from $2$ to $8$ cents. From these two figures, the proposed scheme can achieve a higher AoCR of vehicles and a better payoff of UAV in comparison with the two-tier Q-learning scheme and the greedy scheme. It can be explained as follows. In the two-tier Q-learning scheme, due to the presence of curse of dimensionality resulted from the large state space, UAVs cannot efficiently motivate the high AoCR sharing of vehicles with low payment for enhanced payoffs. In the greedy scheme, as participants greedily seek the optimal AoCR and payment strategies in the Q-learning process, both UAV's payoff and vehicles' AoCR can only stay at the local optimums.

Fig.~\ref{fig:simu6} illustrates the throughput and the block latency in our RescueChain when the block size (i.e., the number of transactions included in a block) increases from $1\times10^3$ to $1\times10^4$.
In this simulation, we set $proposeTimeout\!=\!6$ seconds and $commitTime\!=\!1$ millisecond. As seen in Fig.~\ref{fig:simu6}, our RescueChain system can efficiently cope with thousands of transactions per second with about one-second block latency. Moreover, with the increase of block size, both the blockchain throughput and the block latency increase. The reason is that the higher block size implies more transactions to be processed within a block while it incurs higher block propagation latency.

Fig.~\ref{fig:simu7} compares our RescueChain with other two schemes in terms of the average round to reach consensus, given different ratios of Byzantine nodes.
As seen in Fig.~\ref{fig:simu7}, our RescueChain outperforms the ART scheme and the naive Tendermint scheme in achieving the lowest average round to reach consensus. This is because in the ART scheme, with a reputation decrease (as a punishment) for misbehaving participants, the chance of honest nodes becoming validators can be increased, resulting in a lower average round to reach consensus than the naive Tendermint scheme.
Besides, as the time decay in reputation is absent, previous behaviors occupy a large weight in determining the current reputation, causing a higher risk of suffering spoofing attacks.
Moreover, as the level-2 validators are not considered in the ART scheme in reputation evaluation, the chance of misbehaving participants acting as the leader in blockchain can be increased. Accordingly, the average round to achieve consensus in the ART scheme is relatively larger than that in our RescueChain.

Fig.~\ref{fig:simu8} depicts the average offloading delay in two schemes, where the data size of task $D_{j,k}$ changes from $2$ to $8$ Mbits and three different values of vehicle density $\chi$ are exploited. In the scheme without VFC, UAVs' computation tasks on sensory data are offloaded to the ECVs.
We can see that, compared with the scheme without VFC, our proposed scheme with VFC can efficiently decrease the offloading delay given different vehicular densities and task sizes. This is because the ground vehicles are provisioned with sufficient computation capacities and are closer to UAVs. Moreover, as all UAVs offload their heavy tasks to the remote ECVs, the ECVs' limited computing resource can be fully occupied, resulting in a longer waiting time for computation and increased latency for result delivery.
Besides, as seen in Fig.~\ref{fig:simu8}, given the task size, the average task delay decreases with the increase of vehicle density. The reason is that the higher vehicle density implies more neighboring vehicles for task offloading and the corresponding higher AoCR sharing in computation, causing a decline in the task latency.

Fig.~\ref{fig:simu9} demonstrates the average saved energy of UAV in task offloading given different data sizes of task $D_{j,k}$ and vehicle densities $\chi$. From Fig.~\ref{fig:simu9}, the UAV's saved energy increases with the task size and the vehicle density. It is because with the increase of vehicle density, more computation resources can be contributed to UAVs, thereby saving more limited battery energy of UAVs. Besides, the heavy computing tasks can be effectively offloaded to vehicles under VFC when the task size grows, thereby improving the energy efficiency of UAVs.

\begin{table}[!t]
\setlength{\abovecaptionskip}{-0.02cm}
\caption{Energy consumption of our RescueChain in reaching consensus on a new block, compared with naive Tendermint scheme, Bitcoin, and Ethereum}\label{table4}
\begin{tabular}{|ccc|}
\hline
\multicolumn{1}{|c|}{\multirow{2}{*}{\begin{tabular}[c]{@{}c@{}}\# validators\\ ($Z$)\end{tabular}}} & \multicolumn{2}{c|}{\begin{tabular}[c]{@{}c@{}}Total energy consumption of validators for reaching\\consensus of a block including 1000 transactions (J)\end{tabular}} \\ \cline{2-3}
\multicolumn{1}{|c|}{}                                                                            & \multicolumn{1}{c|}{\textbf{~~~~~~~~~~Proposal~~~~~~~~~~}}                                      & \textbf{Naive Tendermint}                                                                    \\ \hline
\multicolumn{1}{|c|}{$Z$ = 10}                                                                      & \multicolumn{1}{c|}{28.41 J}                                                                       & 28.55 J                                                                             \\ \hline
\multicolumn{1}{|c|}{$Z$ = 50}                                                                      & \multicolumn{1}{c|}{144.02 J}                                                                      & 145.89 J                                                                            \\ \hline
\multicolumn{1}{|c|}{~$Z$ = 100}                                                                     & \multicolumn{1}{c|}{293.04 J}                                                                      & 299.67 J                                                                            \\ \hline
\multicolumn{1}{|c|}{~~$Z$ = 1000}                                                                    & \multicolumn{1}{c|}{3589.95 J}                                                                     & 4413.62 J                                                                           \\ \hline
\multicolumn{3}{|l|}{Average energy consumption per transaction in \textbf{Bitcoin}: $6.48\times 10^9$ J}                                                                                                                                                                                    \\ \hline
\multicolumn{3}{|l|}{Average energy consumption per transaction in \textbf{Ethereum}: $6.44\times 10^8$ J}                                                                                                                                                                                   \\ \hline
\end{tabular}\vspace{-4mm}
\end{table}

Table~\ref{table4} shows the energy efficiency of our RescueChain in the consensus phase in comparison with the naive Tendermint scheme, Bitcoin, and Ethereum. Here, the number of validators in the blockchain varies from $10$ to $1000$, and the block size (i.e., number of transactions included in a block) is set as $1000$. As shown in Table~\ref{table4}, our proposed RescueChain attains a much-improved energy efficiency in reaching consensus than traditional Bitcoin and Ethereum blockchains. {According to \cite{bitnodes,ethernodes}, the numbers of current reachable Bitcoin nodes and Ethereum Mainnet clients are $16169$ and $5683$, respectively. In both Bitcoin and Ethereum, the compute-intensive PoW consensus protocol is leveraged to reach the unambiguous consensus among huge number of miners, thereby consuming dramatically considerable energy in reaching consensus, which is much larger than both naive Tendermint and our RescueChain.}
Besides, the proposed consensus scheme is more energy-efficient than the naive Tendermint scheme, especially when the number of validators is very large (i.e., $1000$). The reason is that thanks to the proposed reputation evaluation mechanism, our RescueChain can enjoy a lower average round to reach consensus on a new block than the naive Tendermint scheme, which is also validated in Fig.~\ref{fig:simu7}. Thereby, our RescueChain can save more energy for validator UAVs in block transmission and vote verification during the consensus process, resulting in improved energy efficiency.

\vspace{-0.1cm}
\subsection{Insights for Implementation}
In this subsection, we discuss the following key components in the practical implementation of the proposed scheme.

\begin{table}[!t]
\setlength{\abovecaptionskip}{-0.02cm}
\caption{Measurements of communication protocols for A2A, A2G/G2A, V2E, and V2V links in UDRNs}\label{table5}
\resizebox{1.0\linewidth}{!}{
\begin{tabular}{|c|c|c|l|c|c|}
\hline
\textbf{Link}            & \textbf{Description} & \textbf{Protocol}                                            & \multicolumn{1}{c|}{\textbf{Data}}                                                            & \multicolumn{1}{c|}{\textbf{Range}} & \multicolumn{1}{c|}{\textbf{Throughput}} \\ \hline
A2A                      & UAV$\leftrightarrow$UAV              & \begin{tabular}[c]{@{}c@{}}Wi-Fi 6 \\ (802.11ax)\end{tabular} & Task data, transactions                                                                       & \textless{}800m                     & \textless{}1.2 GB/s                      \\ \hline
\multirow{2}{*}{\begin{tabular}[c]{@{}l@{}}A2G/\\G2A\end{tabular}} & UAV$\leftrightarrow$vehicle          & \begin{tabular}[c]{@{}c@{}}Wi-Fi 6 \\ (802.11ax)\end{tabular} & \begin{tabular}[c]{@{}l@{}}Task data, transactions,\\ task processing results\end{tabular}                                                                       & \textless{}800m                     & \textless{}1.2 GB/s                      \\ \cline{2-6}
                         & UAV$\leftrightarrow$ECV              & LTE cellular                                                 & \begin{tabular}[c]{@{}l@{}}Task data, transactions,\\scheduling command\end{tabular}     & 1-3 km                               & $\approx$37.5 MB/s                               \\ \hline
V2E                      & vehicle$\leftrightarrow$ECV          & LTE cellular                                                 & \begin{tabular}[c]{@{}l@{}}Transactions, \\scheduling command\end{tabular} & 1-3 km                               & $\approx$37.5 MB/s                               \\ \hline
V2V                      & vehicle$\leftrightarrow$vehicle      & \begin{tabular}[c]{@{}c@{}}WAVE \\ (802.11p)\end{tabular}    & \begin{tabular}[c]{@{}l@{}}Rescue data,\\coordination data \end{tabular}                                                         & \textless{}1 km                      & 3-27 Mb/s                                \\ \hline
\end{tabular}}
\end{table}

\emph{1) Selection of communication protocols.} In the aerial subnetwork, the Wi-Fi 6 (IEEE 802.11ax) protocol that operates in the unlicensed spectrum can be utilized to transmit the sensory data, task data, and transactions among UAVs due to its high data rate (\textless{}1.2 GB/s) and ease of management. In the ground subnetwork, the well-known WAVE (Wireless Access in Vehicular Environments) protocol supported by IEEE 802.11p standard can be available for V2V links among ground rescue vehicles, which utilizes the 5.9 GHz band. Besides, due to the sparse network connections and potential network partitions, if the multi-hop V2V/A2A link is unavailable, the vehicle or UAV can deliver the information in a store-carry-and-forward fashion \cite{7317860}. For the A2G/G2A links between UAVs and ground vehicles, the Wi-Fi module can be used for the transmission of task data, processing results, blockchain transactions, etc. For the ECVs, they can quickly enter part of disaster areas and establish cellular networks with large coverage (about 1-3 km in LTE) and moderate throughput for UAVs and rescue vehicles for rescue coordination (e.g., collision avoidance and position adjustment). Besides, the UAV can offload its computation task to the ECV in its coverage via the cellular module if there are no available ground vehicles or they do not have sufficient computing resources, as a complement to the VFC paradigm. Table~\ref{table5} shows the performance measurements of different communication protocols for A2A, A2G/G2A, vehicle-to-ECV (V2E), and V2V links in UDRNs.

\emph{2) Pre-training DQN models in task offloading.} In DQN-based task offloading, as seen in Figs.~\ref{fig:simu1}--\ref{fig:simu3}, it takes a long time for UAVs and vehicles to learn the optimal offloading strategies, especially in a large-scale network involving a large number of participants. In the real implementation, each UAV and vehicle can pre-train its DQN model by exploiting historical experiences (i.e., previous interaction information) learned from similar scenarios, thereby accelerating learning speed and alleviating energy consumption.

\emph{3) Selection of blockchain parameters.} As shown in Figs.~\ref{fig:simu6}--\ref{fig:simu7} and Table~\ref{table4}, the blockchain parameters such as block size, ratio of Byzantine nodes, and number of validators can have a large effect on the performance of our RescueChain in terms of throughput, block latency, average consensus round, and energy consumption. For example, the higher block size results in higher throughput but also higher block latency; an assumption of low ratio of Byzantine nodes leads to low average consensus rounds but sacrifices consensus security; and more validator UAVs indicates better decentralization but yields higher energy consumption. Thereby, a trade-off needs to be achieved in the practical implementation, where appropriate blockchain parameters should be determined depending on specific scenarios. For example, for the scenario with highly sparse network connections, a low block size is preferred to attain a small block propagation latency; for the scenario with highly constrained UAV battery energy, a low number of UAV validators is preferred to achieve low energy consumption; and for the scenario with few malicious nodes, the small ratio of Byzantine nodes is preferred for faster consensus reaching and higher throughput.

{\emph{4) Cold start issue.} Initially, it is assumed that there exist no less than four connected ECVs in the disaster area to facilitate rescue missions. The genesis block of the blockchain can be created by these ECVs via the consensus process, which mainly includes the initial reputation setting for nodes, the role of the CA, block rewards, creation time, and hash of this block.}
%
\section{Conclusion and Future Work}\label{sec:CONSLUSION}
Our paper focuses on the fundamental issue of provisioning guaranteed security for data dissemination in UDRNs.
We have proposed the RescueChain scheme in disaster sites to address secure data sharing and efficient computation offloading for resource-limited UAVs.
Within this scheme, a lightweight blockchain-based framework has been presented under the collaborative aerial-ground network to immutably trace misbehaving entities and safeguard information sharing. Then, we have devised a reputation-based consensus protocol and a VFC-based off-chain mechanism for improved consensus efficiency and robustness in weakly connected UDRNs.
To stimulate vehicles' collaborative behaviors in VFC-based computation offloading for UAVs, we have formulated the interactions between them as a Stackelberg game and derived the SE of the static Stackelberg game.
Apart from this, in the dynamic Stackelberg game, a novel two-tier DQN-based algorithm has been devised to intelligently and distributively seek the optimal payment strategies for UAVs (tier 1) and computing resource sharing strategies for vehicles (tier 2), without fully knowing the opponent's private parameters and time-varying network parameters.
At last, simulation results have validated the effectiveness of our RescueChain in terms of enhanced consensus and offloading efficiency, reduced task latency and UAVs' energy consumption, and improved payoffs of participants.

For future work, we plan to investigate: 1) the cooperation mechanism among a swarm of UAVs for collaborative sensing, trajectory scheduling, and resource sharing in UDRNs; {2) the coordination of satellites in various earth orbits, UAV swarms, and ground equipments to construct efficient space-air-ground disaster rescue networks; and 3) the vehicle-assisted mobile charging mechanism for efficient UAV battery recharging by equipping wireless energy transfer (WPT) pads on vehicular roofs.}

\begin{appendices}
\section{proof of theorem $4$ }\label{Appendix A}
\begin{IEEEproof}
Note that when $\beta _{i,j,k}=0$, it means that vehicle $i$ will not participate in the offloading process of task $k$ for UAV $j$. Consequently, the corresponding payment of UAV $j$ to vehicle $i$ is zero. As such, $x_{i,j,k}=0$ and $y_{i,j,k}=0$. Here, we only consider the case $\beta _{i,j,k}=1$.
The first order differential of vehicle $i$'s payoff function with respect to $x_{i,j,k}$ is
$\frac{\partial \pi _i\left( x_{i,j,k},y_{i,j,k} \right)}{\partial x_{i,j,k}}=\lambda _py_{i,j,k}-2\lambda _c\psi _ix_{i,j,k}$.
The second order differential for $\pi _i\left( x_{i,j,k},y_{i,j,k} \right)$ with respect to ${q_{i,c}^1}$ is $\frac{\partial ^2\pi _i\left( x_{i,j,k},y_{i,j,k} \right)}{\partial {x_{i,j,k}}^2}=-2\lambda _c\psi _i<0$, which indicates that the function $\frac{\partial \pi _i\left( x_{i,j,k},y_{i,j,k} \right)}{\partial x_{i,j,k}}$ is monotonically decreasing.
The following two cases are considered.

Case $1$: High payment. If the payment of UAV $j$ for task $k$ is high, i.e., $\frac{2\lambda _c\psi _ix_{i}^{\max}}{\lambda _p}\le y_{i,j,k}\le y_{j,k}^{\max}$, we have $\underset{x_{i,j,k}\rightarrow x_{i}^{\max}}{\lim}\frac{\partial \pi _i\left( x_{i,j,k},y_{i,j,k} \right)}{\partial x_{i,j,k}} \ge 0$. As such, $\pi _i\left( x_{i,j,k},y_{i,j,k} \right)$ is a monotonic increasing function. Then, the optimal AoCR strategy of vehicle $i$ is $x_{i,j,k}^{*} = x_{i}^{\max}$.

Case $2$: Low payment. If the payment of UAV $j$ for task $k$ is low, i.e., $0<y_{i,j,k}< \frac{2\lambda _c\psi _ix_{i}^{\max}}{\lambda _p}$, we have $\underset{x_{i,j,k}\rightarrow x_{i}^{\max}}{\lim}\frac{\partial \pi _i\left( x_{i,j,k},y_{i,j,k} \right)}{\partial x_{i,j,k}} < 0$. As such, the utility function of vehicle $i$ is strict convex, and vehicle $i$'s optimal AoCR strategy can be obtained by solving $\frac{\partial \pi _i\left( x_{i,j,k},y_{i,j,k} \right)}{\partial x_{i,j,k}} = 0$, i.e., $x_{i,j,k}^{*} =\frac{\lambda _py_{i,j,k}}{2\lambda _c\psi _i}$.
Theorem~4 is proved.
\end{IEEEproof}

\section{proof of theorem $5$ }\label{Appendix B}
\begin{IEEEproof}
If $\frac{2\lambda _c\psi _ix_{i}^{\max}}{\lambda _p}\le y_{i,j,k}\le y_{j,k}^{\max}$, by substituting $x_{i,j,k}^{*}= x_{i}^{\max}$ into $\pi _j\left( x_{i,j,k},y_{i,j,k} \right)$, the utility function of UAV $j$ can be rewritten as $\pi _j(x_{i,j,k}, y_{i,j,k} ) = -\varpi _p\lambda _py_{i,j,k}x_{i}^{\max} + \rho _j\alpha _{j,k}  \log \left( 1+x_{i}^{\max} \right) -\left( 1\!-\!\varpi _p \right) \Phi _{i,j,k}$.
As the above function is monotonically decreasing with respect to $y_{i,j,k}$, UAV $j$'s optimal payment strategy can be attained as $y_{i,j,k}^{*} = \frac{2\lambda _c\psi _ix_{i}^{\max}}{\lambda _p}$.

If $0 < {p_{i,c}^1} < \frac{2\lambda _c\psi _ix_{i}^{\max}}{\lambda _p}$, by substituting $x_{i,j,k}^{*} = \frac{\lambda _py_{i,j,k}}{2\lambda _c\psi _i}$ into $\pi _j\left( x_{i,j,k},y_{i,j,k} \right)$, UAV $j$'s payoff function can be rewritten as:
\begin{equation}\label{eq:max2-2}
\begin{aligned}
\pi _j( x_{i,j,k},y_{i,j,k} ) &= \rho _j\alpha _{j,k}\log ( 1+\frac{\lambda _py_{i,j,k}}{2\lambda _c\psi _i} )-\frac{\varpi _p( \lambda _p ) ^2}{2\lambda _c \psi _i}( y_{i,j,k} ) ^2 \nonumber \\
& ~~~- \left( 1\!-\!\varpi _p \right) \Phi _{i,j,k}
\end{aligned}
\end{equation}
The first order differential of UAV $j$'s payoff with respect to $y_{i,j,k}$ is $\frac{\partial \pi _j( x_{i,j,k},y_{i,j,k} )}{\partial y_{i,j,k}} \!=\! \frac{\rho _j\alpha _{j,k}\lambda _p}{2\lambda _c\psi _i \!+\! \lambda _py_{i,j,k}} \!-\! \frac{\varpi _p( \lambda _p ) ^2y_{i,j,k}}{\lambda _c\psi _i}$.
The second order differential of UAV $j$'s payoff satisfies $\frac{\partial ^2\pi _j\left( x_{i,j,k},y_{i,j,k} \right)}{\partial y_{i,j,k}^{2}} <0$.
It implies that UAV $j$'s payoff function is strictly convex, and the maximum value can be derived by solving $\frac{\partial \pi _j( x_{i,j,k},y_{i,j,k} )}{\partial y_{i,j,k}} = 0$ with KKT conditions. The optimal payment strategy can be obtained as shown in Eq. (\ref{eq:payment}).
Theorem~5 is proved.
\end{IEEEproof}
\end{appendices}

\bibliographystyle{IEEETran}
\bibliography{ToN-arxiv}

\vspace{-10mm}
\begin{IEEEbiography}[{\includegraphics[width=1in,height=1.25in,clip,keepaspectratio]{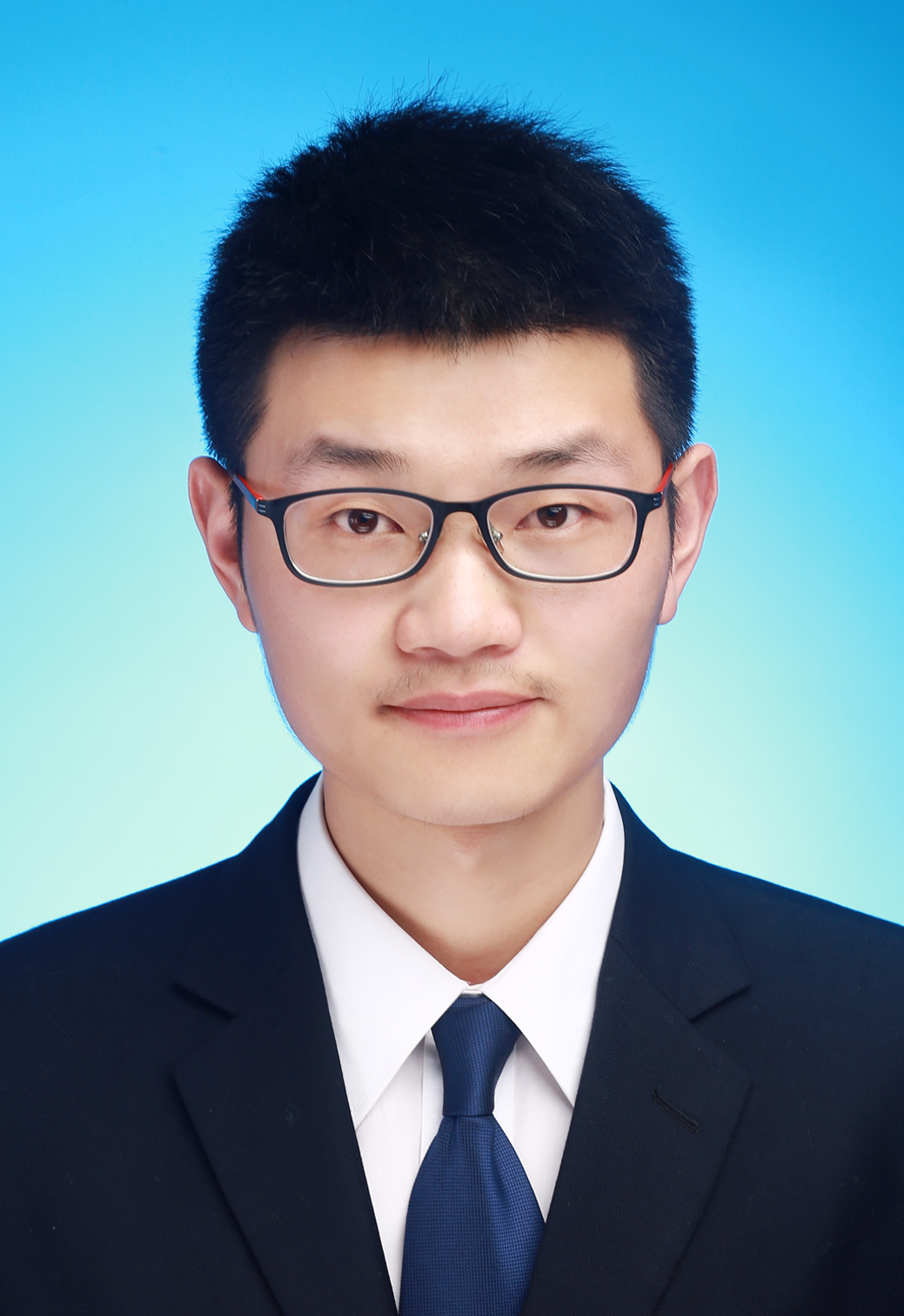}}]{Yuntao Wang}
received the Ph.D degree in Cyberspace Security from Xi'an Jiaotong University, Xi'an, China, in 2022. His research interests include security and privacy in IoT, network games, edge intelligence, and blockchain.
\end{IEEEbiography}\vspace{-5mm}

\begin{IEEEbiography}[{\includegraphics[width=1in,height=1.25in,clip,keepaspectratio]{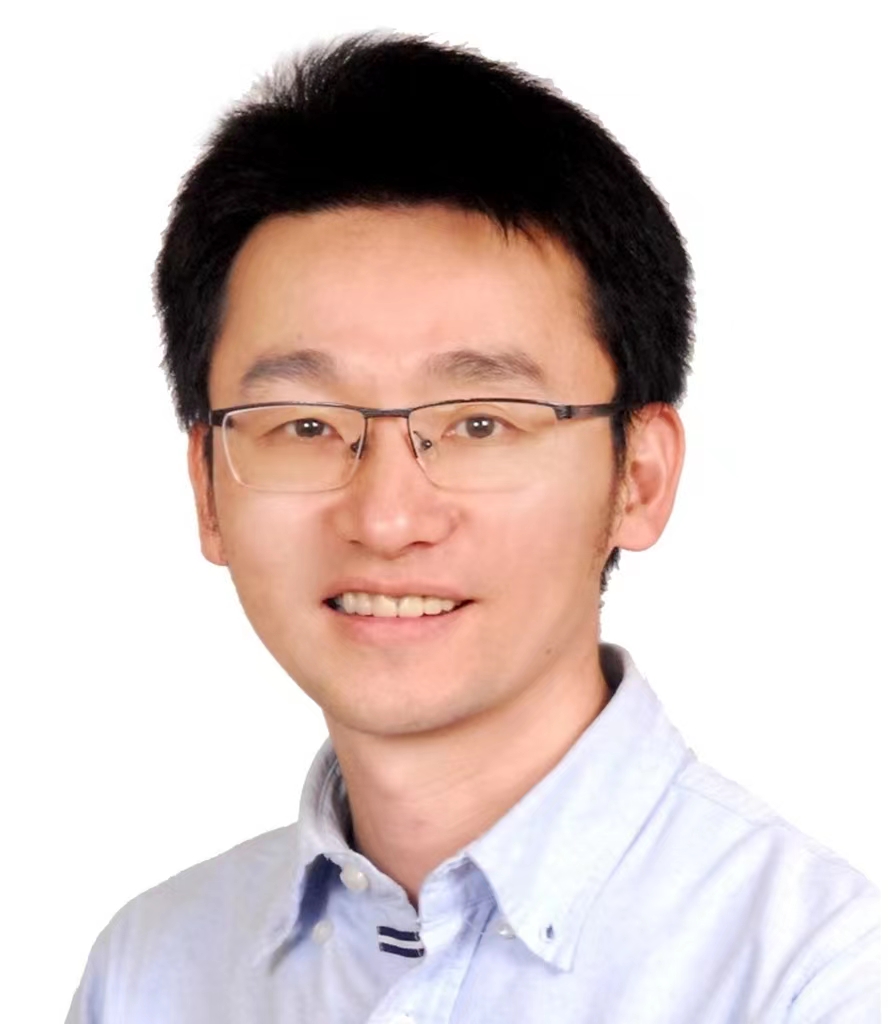}}]{Zhou Su}
has published technical papers, including top journals and top conferences, such as {\scshape IEEE Journal On Selected Areas In Communications}, {\scshape IEEE Transactions On Information Forensics And Security}, {\scshape IEEE Transactions On Dependable And Secure Computing}, {\scshape IEEE Transactions On Mobile Computing}, {\scshape IEEE/ACM Transactions On Networking}, and {\scshape INFOCOM}. His research interests include multimedia communication, wireless communication, and network traffic. Dr. Su received the Best Paper Award of International Conference IEEE ICC2020, IEEE BigdataSE2019, and IEEE CyberSciTech2017. He is an Associate Editor of {\scshape IEEE Internet Of Things Journal}, {\scshape IEEE Open Journal Of Computer Society}, and {\scshape IET Communications}.
\end{IEEEbiography}

\begin{IEEEbiography}[{\includegraphics[width=1in,height=1.25in,clip,keepaspectratio]{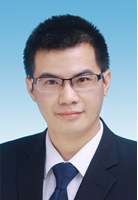}}]{Qichao Xu}
received the Ph.D. degree from the School of Mechatronic Engineering and Automation, Shanghai University, Shanghai, China, in 2019. He is currently an Assistant Professor with Shanghai University. His research interests include wireless network architecture and vehicular networks.
\end{IEEEbiography}

\begin{IEEEbiography}[{\includegraphics[width=1in,height=1.25in,clip,keepaspectratio]{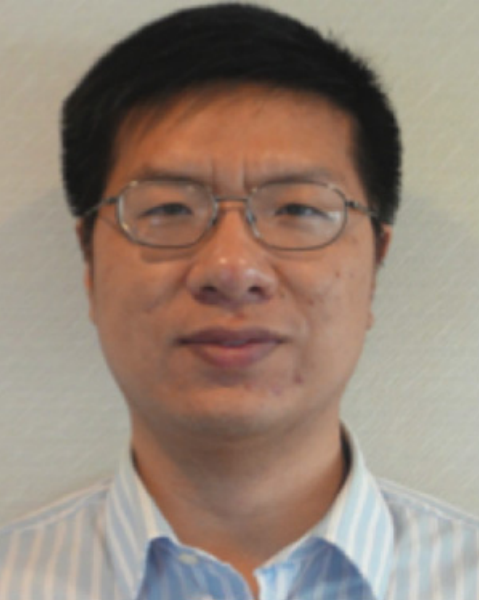}}]{Ruidong Li}
received the D.Eng. degree from the University of Tsukuba in 2008. He is currently an Associate Professor with the College of Science and Engineering, Kanazawa University, Japan. His current research interests include future networks, big data networking, blockchain, the Internet of Things, and network security. He is the Secretary of IEEE ComSoC Internet Technical Committee and the Founder and Chair of the IEEE SIG on big data intelligent networking and IEEE SIG on intelligent Internet edge.
He is a guest editor of prestigious journals, such as IEEE Communications Magazine, IEEE Network Magazine, and IEEE Transactions on Network Science and Engineering.
\end{IEEEbiography}

\begin{IEEEbiography}[{\includegraphics[width=1in,height=1.25in,clip,keepaspectratio]{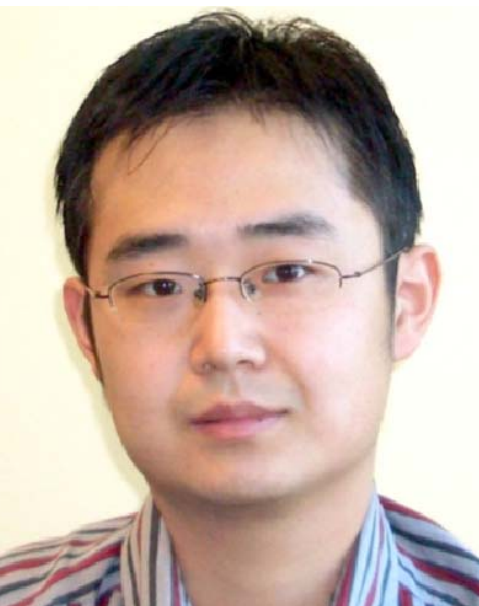}}]{Tom H. Luan}
received the Ph.D. degree from the University of Waterloo, Canada, in 2012. He is currently a Professor with the School of Cyber Science and Engineering, Xi'an Jiaotong University, China. He has authored/coauthored more than 90 journal articles and 30 technical articles in conference proceedings. His research mainly focuses on content distribution and media streaming in vehicular ad hoc networks and peer-to-peer networking and the protocol design and performance evaluation of wireless cloud computing and edge computing. He served as a TPC Member for IEEE Globecom, ICC, and PIMRC.
\end{IEEEbiography}

\begin{IEEEbiography}[{\includegraphics[width=1in,height=1.25in,clip,keepaspectratio]{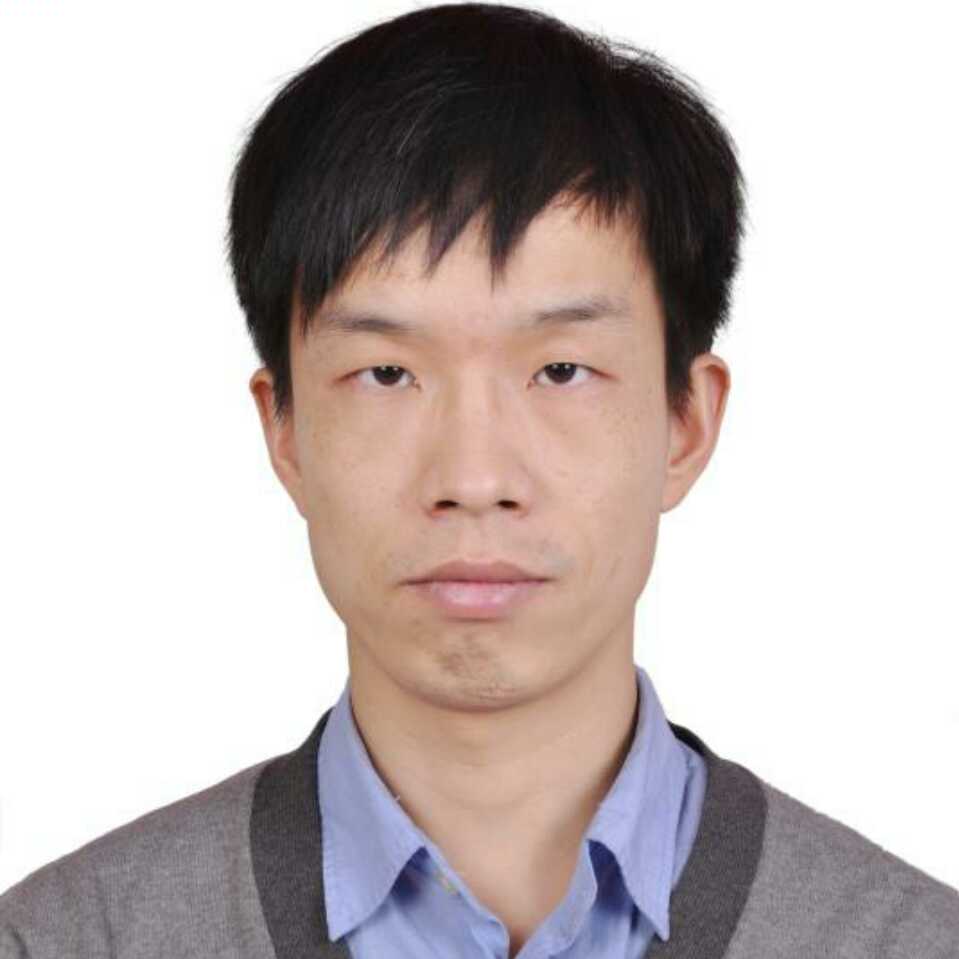}}]{Pinghui Wang}
received the BS and PhD degrees in information engineering from Xi'an Jiaotong University, Xi’an, China, in 2006 and 2012, respectively. He is currently a professor with the MOE Key Laboratory for Intelligent Networks and Network Security, Xi'an Jiaotong University, China. His research interests include Internet traffic measurement and modeling, traffic classification, abnormal detection, and online social network measurement.
\end{IEEEbiography}

\end{document}